\documentclass[aps]{revtex4}
\usepackage{eurosym}
\usepackage{amsfonts}
\usepackage{amsmath}
\usepackage{amssymb,epsf}
\usepackage{color}
\usepackage{graphicx}
\usepackage{float}
\usepackage{caption}
\usepackage{subfig}
\usepackage{epstopdf}

\begin{document}

\title{Charged 4D Einstein-Gauss-Bonnet-AdS Black Holes: \\
Shadow, Energy Emission, Deflection Angle and Heat Engine}
\author{B. Eslam Panah$^{1,2,3,4,5}$\footnote{
email address: beslampanah@shirazu.ac.ir}, Kh. Jafarzade$^{1}$\footnote{
email address: khadije.jafarzade@gmail.com}, and S. H. Hendi$^{5,6}$ \footnote{%
email address: hendi@shirazu.ac.ir}}
\affiliation{$^{1}$ Sciences Faculty, Department of Physics, University of Mazandaran, P.
O. Box 47415-416, Babolsar, Iran\\
$^{2}$ Research Institute for Astronomy and Astrophysics of Maragha (RIAAM),
Maragha, Iran\\
$^{3}$ ICRANet-Babolsart, University of Mazandaran, P. O. Box 47415-416,
Babolsar, Iran\\
$^{4}$ National Elites Foundation of Iran, Tehran, Iran\\
$^{5}$ Department of Physics, School of Science, Shiraz University, Shiraz
71454, Iran\\
$^{6}$ Biruni Observatory, School of Science, Shiraz University, Shiraz
71454, Iran}

\begin{abstract}
Recently, there has been a surge of interest in the 4D Einstein-Gauss-Bonnet
(4D EGB) gravity theory which bypasses the Lovelock theorem and avoids
Ostrogradsky's instability. Such a novel theory has nontrivial dynamics and
presents several predictions for cosmology and black hole physics. Motivated
by recent astrophysical observations and the importance of anti-de Sitter
spacetime, we investigate shadow geometrical shapes and deflection angle of
light from the charged AdS black holes in 4D EGB gravity theory. We explore
the shadow behaviors and photon sphere around such black holes, and inspect
the effect of different parameters on them. Then, we present a study
regarding the energy emission rate of such black holes and analyze the
significant role of the Gauss-Bonnet (GB) coupling constant in the radiation process. Then,
we perform a discussion of holographic heat engines of charged 4D EGB-AdS
black holes by obtaining the efficiency of a rectangular engine cycle.
Finally, by comparing heat engine efficiency with the Carnot efficiency, we
indicate that the ratio $\frac{\eta }{\eta _{c}}$ is always less than one
which is consistent with the thermodynamic second law.
\end{abstract}

\maketitle

\section{Introduction}

To understand the low-energy limit of string theory a noteworthy number of
attempts in higher dimensions theory of gravity have been done (it is
notable that string theory plays a major role as a candidate for the quantum
gravity and also the unification of all interactions. This theory of gravity
needs to higher dimensions for its mathematical consistency).
Einstein-Gauss-Bonnet (EGB) gravity is an important higher dimensional
generalization of Einstein gravity which first was suggested by Lanczos in
1938 \cite{Lanczos}, and then rediscovered by Lovelock in 1971 \cite%
{Lovelock}. EGB gravity includes string theory inspired corrections to the
Einstein-Hilbert action and admits Einstein gravity as a particular case
\cite{Gross}. The study of EGB gravity becomes very important since it
provides a broader set up to explore a lot of conceptual issues related to
gravity. The other interesting properties of EGB gravity are including: i)
it encompasses Einstein gravity as a special case, ii) this theory of
gravity similar to Einstein gravity enjoys only first and second order
derivatives of the metric function in the field equations. iii) it may lead
to the modified Renyi entropy \cite{Pastras}. iv) EGB gravity is free from
the ghosts. v) regarding the AdS/CFT correspondence, it was shown that
considering the EGB theory of gravity will modify entropy, electrical
conductivity, shear viscosity and also thermal conductivity (see Ref. \cite%
{Hu}, for more details). vi) from the cosmological point of view, EGB
gravity can yield a viable inflationary era, and also can describe
successfully the late-time acceleration era, \cite%
{Kanti,Cognola,Satoh,ZGuo,Capozziello,NojiriII,Jiang,Bamba,NojiriI,Nozari,OikonomouOI,Rashidi}%
. See Refs. \cite%
{Oikonomou,EGB0,EGBI,EGBII,EGBIII,OdintsovOE2,OdintsovOF,EGBIV,EGBV,EGBVI,EGBVII,EGBVIII,EGBIX}%
, for more properties of EGB gravity.

It is notable that the case $4$-dimensional of EGB gravity is special
because the Euler-Gauss-Bonnet term becomes a topological invariant in which
does not contribute to the equations of motion or the gravitational
dynamics. In order to make the EGB combination dynamical in four dimensions,
we can use both the nonminimal coupling to the dilaton field \cite%
{DilatonI,DilatonII,DilatonIII,DilatonIV,DilatonV,DilatonVI}, and conformal
anomaly \cite{Cai4I,Cai4II}. Recently, Glavan and Lin in Ref. \cite{Glavan},
have introduced a general covariant modified theory of gravity in $4$%
-spacetime dimensions $\left( D=4\right)$ which propagates only the massless
graviton and also bypasses the Lovelock's theorem (according to the
Lovelock's theorem \cite{Lanczos,Lovelock}, Einstein gravity (with the
cosmological constant) is the unique theory of gravity if we respect to
several conditions: spacetime is $4$-dimensional, metricity, diffeomorphism
invariance, and also the second order equations of motion).\ Indeed, this
theory is formulated in higher dimensions (more than $4$-dimensional
spacetime, $D>4$)\ and its action consists of the Einstein-Hilbert term with
a cosmological constant, and also the GB coupling has been
rescaled as $\alpha /(D-4)$. The $4$-dimensional theory is defined as the
limit\ $D\rightarrow 4$. In this limit, GB invariant gives rise to
nontrivial contributions to gravitational dynamics while preserving the
number of graviton degrees of freedom and also being free from the
Ostrogradsky instability. This new theory is called $4D$ EGB gravity (see
Ref. \cite{Glavan}, for more details). However, some serious questions are
being mentioned on the overall acceptability of the limiting process, the
validity of the equation in $4D$ as well as the absence of proper action and
a consistent theory in $4D$ (see Refs. \cite%
{PobI,PobII,PobIII,PobIV,PobV,PobVI,PobVII}, for more details). In
particular, it is fair to say that the jury is out on this issue, and we
have to wait for some time before the air is cleared.

$4D$ EGB gravity enjoys the existence of several attractive properties. For
example, it might resolve some singularity issues. Indeed, by considering $%
4D $ EGB gravity the static and spherically symmetric black holes the
gravitational force is repulsive at small distances and thus an infalling
particle never reaches the singularity. In addition, static and spherically
symmetric black hole solutions in this new theory differ from the well-known
Schwarzschild black hole in Einstein gravity. Compact objects and their
properties in $4D$ EGB gravity have been studied by many authors. Some of
these works are; quasinormal modes, stability, strong cosmic censorship and
shadows of a black hole \cite{Konoplya,Mishra,RoyCh}, the innermost stable
circular orbit and shadow \cite{GuoLi}, rotating black hole shadow \cite%
{WeiLiu}, Bardeen black holes \cite{DVSingh,AKumar}, thermodynamics, phase
transition and Joule Thomson expansion of (un)charged AdS black hole \cite%
{Hegde,WeiLiuII,WangSL}, rotating black holes \cite{KumarG,GhoshM},
Born-Infeld black holes \cite{KYang}, relativistic stars solution \cite%
{Doneva}, spinning test particle orbiting around a static spherically
symmetric black hole \cite{ZhangWL}, bending of light in $4D$ EGB black
holes by Rindler-Ishak method \cite{FardfSI}, thermodynamics and $P-V$
criticality of AdS black hole \cite{Singh,DVSingh2}, the eikonal
gravitational instability of asymptotically flat and (A)dS black holes\ \cite%
{KonoplyaZ}, greybody factor and power spectra of the Hawking radiation \cite%
{ZhangLG}, stability of the Einstein static universe in this theory \cite%
{LiWYI}, charged particle and epicyclic motions around $4D$ EGB black hole
immersed in an external magnetic field \cite{Shaymatov}, thermodynamic
geometry of AdS black hole \cite{Mansoori}, gravitational lensing by black
holes \cite{Islam}, Hayward black holes \cite{KumarSII}, superradiance and
stability of charged $4D$ EGB black holes \cite{ZhangZhang}, and thin
accretion disk around of black hole \cite{LiuZW}.

The Shadow of black hole is an image of the photon sphere. The Schwarzschild
black hole shadow has a perfect circle shape which is at $r=3M$ \cite%
{Synge,Perlick,Cunha}, but for a Kerr black hole is not a perfect circle
\cite{BardeenBH3}. In $1973$, Bardeen has studied the shadow of Kerr black
holes \cite{BardeenBH3}. In this regard, some people have studied the black
hole shadow in Refs. \cite%
{ShadowO,ShadowI,ShadowII,ShadowIII,ShadowIV,ShadowV,Jafarzade2}. In 2019,
direct observation from the first image of the shadow of a black hole in $%
M87^{\ast} $ galaxy announced by the Event Horizon Telescope (EHT)
international collaboration \cite{AkiyamaI,AkiyamaII}. This observation
attracted more attention to study of black hole shadow \cite%
{KumarG,ShadowVI,ShadowVII,ShadowVIII,ShadowIX,ShadowX,ShadowXI,ShadowXII,ShadowXIII,ShadowXV,ShadowXVI,ShadowXVII,ShadowXVIII,ShadowXIX,ShadowXX,ShadowXXI,ShadowXXII,ShadowXXIII}.

This paper is organized as follows. In the next section, we first give a
brief review of the charged AdS black hole solutions in $4D$ EGB gravity
since it is the base of our present work. Then, we investigate the shadow
behavior of such black holes and explore the effect of black hole (BH) parameters on the size of photon orbits and spherical shadow. Studying the associated
energy emission rate, we analyze the effective role of BH parameters on the
emission of particles around such black holes. Also, we discuss the
influence of different parameters on the light deflection around this kind
of black holes. In Sec. III, we present a study of thermodynamic features of
these black holes in the context of holographic heat engine and examine how
the parameters affect the efficiency of black hole heat engine. Comparing
the engine efficiency to Carnot efficiency, we provide consistent results
with the thermodynamic second law.

\section{ Charged AdS black hole in 4D EGB gravity}

In this section, we first review the charged AdS black hole
solution in the mentioned gravity. Then, we investigate the shadow
behavior of the black hole and examine the influence of black hole
parameters on this optical quantity. Also, we explore the role of
these parameters on the energy emission rate and the light
deflection around this black hole.

At first, we construct the action of
$\mathbf{D-}$dimensional EGB gravity coupled with the Maxwell
field in such a way that the GB coupling constant $\alpha$ is
rescaled with $\alpha /(D-4)$ as 
\begin{equation}
\mathcal{I}=\frac{1}{16\pi }\int d^{D}x\sqrt{-g}\left( R-2\Lambda +\frac{%
\alpha }{D-4}\mathcal{L}_{GB}-F_{\mu \nu }F^{\mu \nu }\right) ,  \label{acI}
\end{equation}%
where $\Lambda$ is the cosmological constant and $F_{\mu
\nu }=\partial _{\mu }A_{\nu }-\partial _{\nu }A_{\mu }$ indicates
the Faraday tensor ($A_{\mu }$ is the gauge potential). In the
above action, $\mathcal{L}_{GB}$ is the Lagrangian of GB gravity with the
following explicit form
\begin{equation}
\mathcal{L}_{GB}=R_{\mu \nu \lambda \rho }R^{\mu \nu \lambda \rho }-4R_{\mu
\nu }R^{\mu \nu }+R^{2}.
\end{equation}

By variation of the action (\ref{acI}) with respect to the
metric ($g_{\mu \nu }$) and Faraday tensor ($F_{\mu \nu }$), one
can obtain the following field equations 
\begin{eqnarray}
G_{\mu \nu }-\Lambda g_{\mu \nu }+\frac{\alpha }{D-4}H_{\mu \nu } &=&T_{\mu
\nu },  \label{FieldI} \\
&&  \notag \\
\nabla _{\mu }F^{\mu \nu } &=&0,  \label{FieldII}
\end{eqnarray}%
where $G_{\mu \nu }$ is the Einstein tensor while $H_{\mu
\nu } $ and $T_{\mu \nu }$ are
\begin{eqnarray}
H_{\mu \nu } &=&2\left( RR_{\mu \nu }-2R_{\mu \sigma }R_{\nu }^{\sigma
}-2R_{\mu \sigma \nu \rho }R^{\sigma \rho }-R_{\mu \sigma \rho \beta }R_{~\
\ \nu }^{\sigma \rho \beta }\right) -\frac{1}{2}g_{\mu \nu }\mathcal{L}_{GB},
\\
&&  \notag \\
T_{\mu \nu } &=&2F_{\mu }^{~\ \lambda }F_{\nu \lambda }-\frac{1}{2}g_{\mu
\nu }F^{\alpha \beta }F_{\alpha \beta }.
\end{eqnarray}

Here, we are going to investigate black hole solutions
described with a $4-$dimensional static spherically symmetric
metric in the following form
\begin{equation}
ds^{2}=-f(r) dt^{2}+\frac{1}{f(r)}
dr^{2}+r^{2}(d\theta^{2}+sin^2\theta\; d\phi^2), \label{Metric}
\end{equation}

Regarding this metric with a consistent gauge potential
ansatz $A_{\mu }=h(r)\delta _{\mu }^{0}$, one can use the Maxwell
equation (\ref{FieldII}) to obtain the electric field as
$h(r)=-\frac{Q}{r}$ \cite{Hegde,Fernandes1}.

Now, we should solve the gravitational field equation
(\ref{FieldII}) for unknown $f(r)$ in arbitrary $D-$dimensions.
Then, by setting the limit $D\rightarrow 4$, the static
spherically symmetric charged AdS black hole solution in $4D$ EGB
gravity is given by \cite{Hegde,Fernandes1}
\begin{equation}
f(r)=1+\frac{r^{2}}{2\alpha }\left( 1\pm \sqrt{1+4\alpha \left( \frac{2M}{%
r^{3}}-\frac{Q^{2}}{r^{4}}+\frac{\Lambda }{3}\right) }\right) ,
\label{Eqmetric1}
\end{equation}%
where $M$ and $Q$ can be identified, respectively, as the mass and charge
parameters of the black hole. Equation (\ref{Eqmetric1}) corresponds to two
branches of solutions depending on the choice of $\pm $. Since $+$ve branch
does not lead to a physically meaningful solution \cite{BDghost}, we will
limit our discussions to $-$ve branch of the solution (\ref{Eqmetric1}).

\subsection{Photon sphere and shadow}

Here, we would like to explore the shadow of this black hole and investigate
the effect of different parameters on the size of photon orbits and
spherical shadow. To do so, we employ the Hamilton-Jacobi method for a
photon in the black hole spacetime as \cite{Carter,Decanini}
\begin{equation}
\frac{\partial S}{\partial \sigma }+H=0,
\end{equation}%
where $S$ and $\sigma $ are, respectively, the Jacobi action and affine
parameter along the geodesics. A massless photon moving in the static
spherically symmetric spacetime can be controlled by the following
Hamiltonian
\begin{equation}
H=\frac{1}{2}g^{ij}p_{i}p_{j}=0.  \label{EqHamiltonian}
\end{equation}

Due to the spherically symmetric property of the black hole, we can consider
photons moving on the equatorial plane with $\theta =\frac{\pi }{2}$. So,
Eq. (\ref{EqHamiltonian}) reduces to
\begin{equation}
\frac{1}{2}\left[ -\frac{p_{t}^{2}}{f(r)}+f(r)p_{r}^{2}+\frac{p_{\phi }^{2}}{%
r^{2}}\right] =0.  \label{EqNHa}
\end{equation}

Since the Hamiltonian does not depend explicitly on the coordinates $t$ and $%
\phi $, one can define two constants of motion as
\begin{equation}
p_{t}\equiv \frac{\partial H}{\partial \dot{t}}=-E~~~~\&~~~~p_{\phi }\equiv
\frac{\partial H}{\partial \dot{\phi}}=L,  \label{Eqenergy}
\end{equation}%
where the quantities $E$ and $L$ are interpreted as the energy and angular
momentum of the photon, respectively. Using the Hamiltonian formalism, the
equations of motion can be obtained as
\begin{eqnarray}
\dot{t} &=&\frac{\partial H}{\partial p_{t}}=-\frac{p_{t}}{f(r)},  \notag \\
&&  \notag \\
\dot{r} &=&\frac{\partial H}{\partial p_{r}}=p_{r}f(r),  \notag \\
&&  \notag \\
\dot{\phi} &=&\frac{\partial H}{\partial p_{\phi }}=-\frac{p_{\phi }}{r^{2}},
\label{Eqmotion}
\end{eqnarray}%
where the dot is derivative with respect to the affine parameter $\sigma $.
These equations provide a complete description of the dynamics with the
effective potential as
\begin{equation}
\dot{r}^{2}+V_{eff}(r)=0~~~\Longrightarrow ~~~V_{eff}(r)=f(r)\left[ \frac{%
L^{2}}{r^{2}}-\frac{E^{2}}{f(r)}\right] .  \label{Eqpotential}
\end{equation}

It is worthwhile to mention that the photon orbits are circular and unstable
associated to the maximum value of the effective potential. Radius of the
photon sphere can be obtained from the following conditions
\begin{equation}
V_{eff}(r)\Bigg\vert_{r=r_{p}}=0,~~\&~~~\frac{\partial V_{eff}(r)}{\partial r%
}\Bigg\vert_{r=r_{p}}=0,  \label{Eqcondition}
\end{equation}%
which results into the following equation%
\begin{equation}
(3+4\alpha \Lambda )r_{p}^{4}-27M^{2}r_{p}^{2}+12Mr_{p}(2\alpha
+3Q^{2})-12Q^{2}(\alpha +Q^{2})=0,  \label{Eqshadow1}
\end{equation}%
where $r_{P}$ is the photon sphere radius.

Since Eq. (\ref{Eqshadow1}) is complicated to determine $r_{p}$
analytically, we employ a numerical method for solving this equation. The
event horizon ($r_{e}$ in which obtain by $f(r)\Bigg\vert_{r=r_{e}}=0$) and
the radius of unstable photon sphere (larger photon orbit) for some
parameters are listed in table \ref{table1}. As we see, by considering fixed
values for the GB parameter and the cosmological constant, increasing the
electric charge leads to decreasing $r_{e}$ and $r_{p}$. A similar
explanation can be used regarding the GB parameter with keeping the
cosmological constant and electric charge as constant. A significant point
about these two parameters is that as $Q$ and $\alpha $ increase, some
constraints are imposed on these parameters due to the imaginary event
horizon. Studying $\Lambda $ effect, we observe that as $\Lambda $ increases
from $-0.2$ to $-0.01$, $r_{e}$ ($r_{P}$) increases (decreases).
\begin{table}[tbh]
\caption{Event horizon and photon sphere radius for variation of electric
charge, GB parameter and cosmological constant for $M=1$. }
\label{table1}\centering
\begin{tabular}{ccccccc}
{\footnotesize $Q$ \hspace{0.3cm}} & \hspace{0.3cm}$0.02$ \hspace{0.3cm} &
\hspace{0.3cm} $0.2$\hspace{0.3cm} & \hspace{0.3cm} $0.5$\hspace{0.3cm} &
\hspace{0.3cm}$0.8$\hspace{0.3cm} & \hspace{0.3cm}$0.9 $ \hspace{0.3cm} &
\hspace{0.3cm} \\ \hline\hline
$r_e (\alpha=0.2 $, $\Lambda=-0.02 $) & $1.84 $ & $1.82 $ & $1.70$ & $1.36$
& $0.9+0.12I$ &  \\ \hline
$r_p (\alpha=0.2 $, $\Lambda=-0.02 $) & $2.91 $ & $2.88 $ & $2.72$ & $2.33$
& $2.08$ & {\ \hspace{0.3cm}} \\ \hline
&  &  &  &  &  &  \\
{\footnotesize $\alpha$ \hspace{0.3cm}} & \hspace{0.3cm}$0.03$ \hspace{0.3cm}
& \hspace{0.3cm} $0.3$\hspace{0.3cm} & \hspace{0.3cm} $0.6$\hspace{0.3cm} &
\hspace{0.3cm}$0.9$\hspace{0.3cm} & \hspace{0.3cm}$1 $ &  \\ \hline\hline
$r_e (Q=0.2 $, $\Lambda=-0.02 $) & $1.91 $ & $1.77$ & $1.56$ & $1.21$ & $%
0.9+0.2I $ &  \\ \hline
$r_p (Q=0.2 $, $\Lambda=-0.02 $) & $2.96 $ & $2.83$ & $2.67$ & $2.45$ & $%
2.36 $ & {\ \hspace{0.3cm}} \\ \hline
&  &  &  &  &  &  \\
{\footnotesize $\Lambda$ \hspace{0.3cm}} & \hspace{0.3cm}$-0.01$ \hspace{%
0.3cm} & \hspace{0.3cm} $-0.05$\hspace{0.3cm} & \hspace{0.3cm} $-0.1$\hspace{%
0.3cm} & \hspace{0.3cm}$-0.15$\hspace{0.3cm} & \hspace{0.3cm} $-0.2 $
\hspace{0.3cm} & \hspace{0.3cm} \\ \hline\hline
$r_e$ ($\alpha=Q=0.2$) & $1.84$ & $1.77$ & $1.69$ & $1.63$ & $1.58$ &  \\
\hline
$r_p$ ($\alpha=Q=0.2$) & $2.88$ & $2.89$ & $2.91$ & $2.93$ & $2.96$ &  \\
\hline
\end{tabular}%
\end{table}

The orbit equation for the photon is obtained as
\begin{equation}
\frac{dr}{d\phi}=\frac{\dot{r}}{\dot{\phi}}=\frac{r^{2}f(r)}{L}p_{r}.
\label{Eqorbit}
\end{equation}

Using Eq. (\ref{Eqmotion}), one finds
\begin{equation}
\frac{dr}{d\phi }=\pm r\sqrt{f(r)\left[ \frac{r^{2}E^{2}}{f(r)L^{2}}-1\right]
}.  \label{EqNorbit}
\end{equation}

The turning point of the photon orbit is expressed by the constraint $\frac{%
dr}{d\phi}\Big\vert_{r=R}=0 $. Then we have
\begin{equation}
\frac{dr}{d\phi}=\pm r\sqrt{f(r)\left[\frac{r^{2}f(R)}{R^{2}f(r)} -1\right] }%
.  \label{EqTp}
\end{equation}

Considering a light ray sending from a static observer placed at $r_{0} $
and transmitting into the past with an angle $\vartheta$ with respect to the
radial direction, one has \cite{M.Zhang,Belhaj}
\begin{equation}
\cot \vartheta =\frac{\sqrt{g_{rr}}}{g_{\phi\phi}}\frac{dr}{d\phi}\Bigg\vert%
_{r=r_{0}}.  \label{Eqangle}
\end{equation}

Hence, one can obtain the shadow radius of the black hole as
\begin{equation}
r_{s}=r_{0}\sin \vartheta =R\sqrt{\frac{f(r_{0})}{f(R)}}\Bigg\vert_{R=r_{p}}.
\label{Eqshadow}
\end{equation}

\begin{figure}[tbh]
\centering
\subfloat[$ \alpha=0.2 $ and $ \Lambda =-0.02 $]{
        \includegraphics[width=0.32\textwidth]{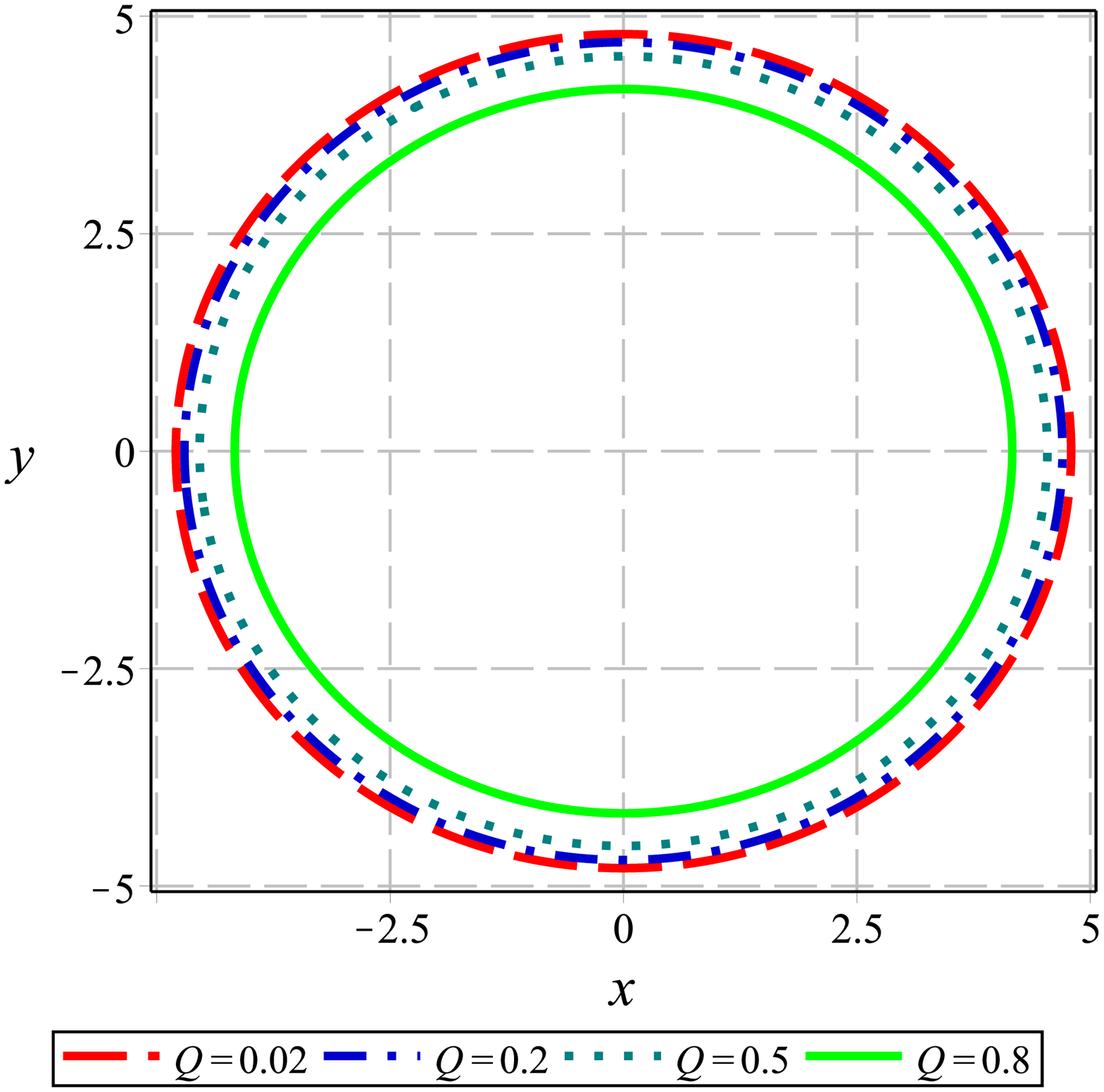}}
\subfloat[$Q=0.2$ and $ \Lambda =-0.02 $]{
        \includegraphics[width=0.32\textwidth]{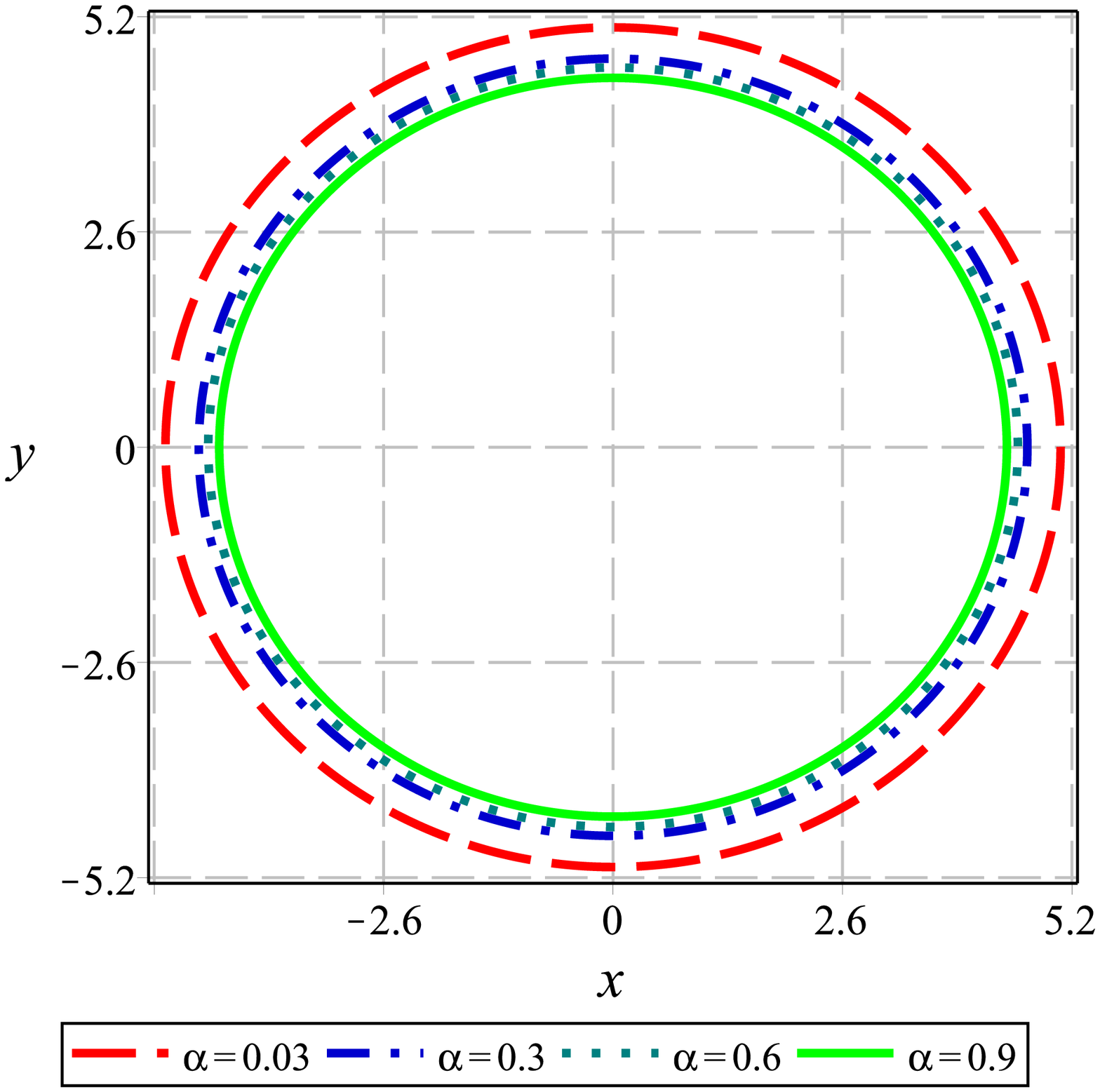}}
\subfloat[$ Q=\alpha=0.2 $ ]{
        \includegraphics[width=0.315\textwidth]{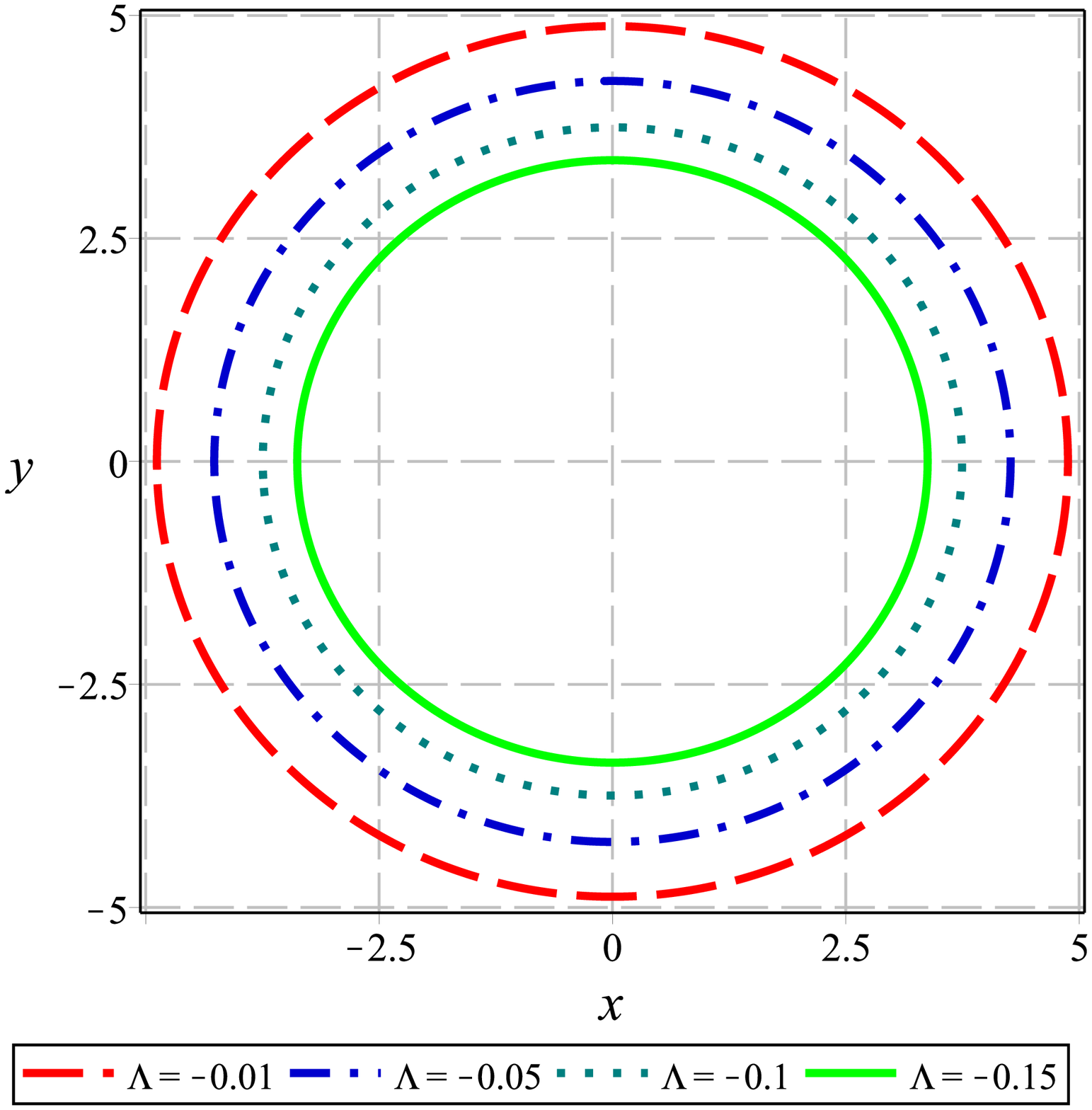}}\newline
\caption{Black hole shadow in the Celestial plane $(x-y)$ for $M=1$.}
\label{Fig1}
\end{figure}

The apparent shape of a shadow is obtained by using the celestial
coordinates $x$ and $y$ which are defined as \cite{Vazquez,RShaikh}
\begin{eqnarray}
x &=&\lim_{r_{0}\longrightarrow \infty }\left( -r_{0}^{2}\sin \theta _{0}%
\frac{d\phi }{dr}\Big\vert_{(r_{0},\theta _{0})}\right) ,  \notag \\
&&  \notag \\
y &=&\lim_{r_{0}\longrightarrow \infty }\left( r_{0}^{2}\frac{d\theta }{dr}%
\Big\vert_{(r_{0},\theta _{0})}\right) ,
\end{eqnarray}%
where $(r_{0},\theta _{0})$ are the position coordinates of the observer. To
investigate the effect of electric charge, GB parameter and the cosmological
constant on the size of black hole shadow, we have plotted Fig. \ref{Fig1}%
. From this figure, one can find that the size of circular shape of the
shadow shrinks with increasing $Q$ and $\alpha $. As we see from Fig. \ref%
{Fig1}a, the effect of electric charge would be significant for its large
values, whereas GB coupling constant will have notable impact for its small
values (see Fig. \ref{Fig1}b).

In Fig. \ref{Fig1}c, we plot the black hole shadow for different values of
the cosmological constant. We observe that increasing $\Lambda $ makes the
increasing of the shadow radius. Comparing Fig. \ref{Fig1}c to Figs. \ref%
{Fig1}a and \ref{Fig1}b, it is evident that $\Lambda $ has a stronger effect
on the shadow size than $Q$ and $\alpha $.

\subsection{Energy emission rate}

Now, we would like to study the associated energy emission rate. It was
shown that for a far distant observer, the absorption cross-section
approaches to the black hole shadow \cite{WWei,Belhaj1,Belhaj2}. In
general, the absorption cross-section oscillates around a limiting constant
value $\sigma _{lim}$ at very high energy. It was found that $\sigma _{lim}$
is approximately equal to the area of the photon sphere $\left( \sigma
_{lim}\approx \pi r_{s}^{2}\right) $ which provides the energy emission rate
expression given by
\begin{equation}
\frac{d^{2}E(\varpi )}{dtd\varpi }=\frac{2\pi ^{3}\varpi ^{3}r_{s}^{2}}{e^{%
\frac{\varpi }{T}}-1},  \label{Eqemission}
\end{equation}%
where $\varpi $ is the emission frequency. $T$ denotes the Hawking
temperature. Hawking temperature related to surface gravity $\kappa $ on
event horizon is calculated as
\begin{equation}
T=\frac{\kappa }{2\pi }=\frac{f^{\prime }\left( r_{e}\right) }{4\pi }=\frac{%
r_{e}^{2}-Q^{2}-\alpha }{4\pi r_{e}\left( r_{e}^{2}+2\alpha \right) }-\frac{%
\Lambda r_{e}^{3}}{4\pi \left( r_{e}^{2}+2\alpha \right) }.  \label{EqT-H}
\end{equation}

\begin{figure}[tbh]
\centering
\subfloat[ $ \alpha=0.2 $ and $ \Lambda =-0.02 $]{
        \includegraphics[width=0.32\textwidth]{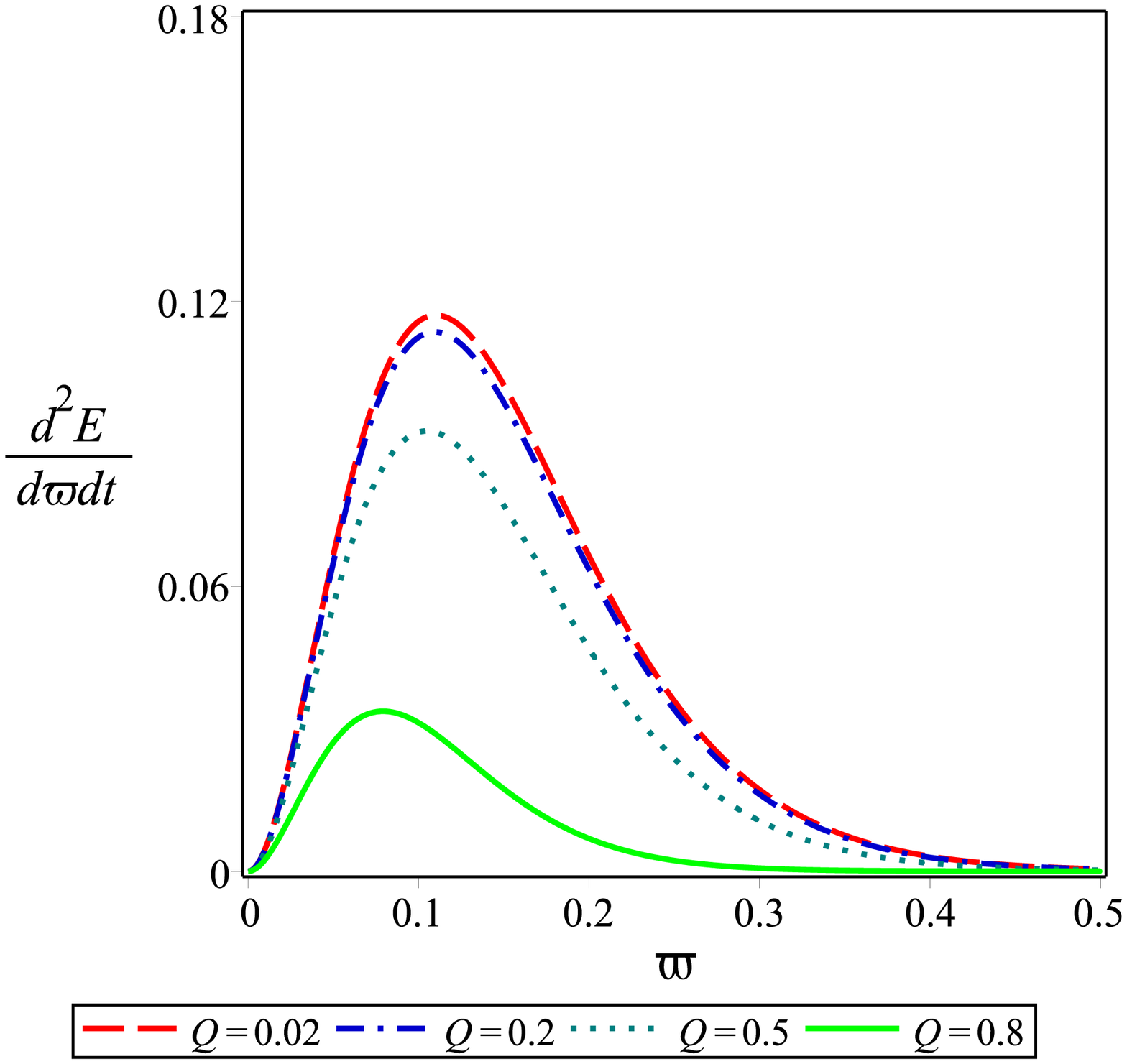}}
\subfloat[$Q=0.2$ and $ \Lambda =-0.02 $]{
        \includegraphics[width=0.32\textwidth]{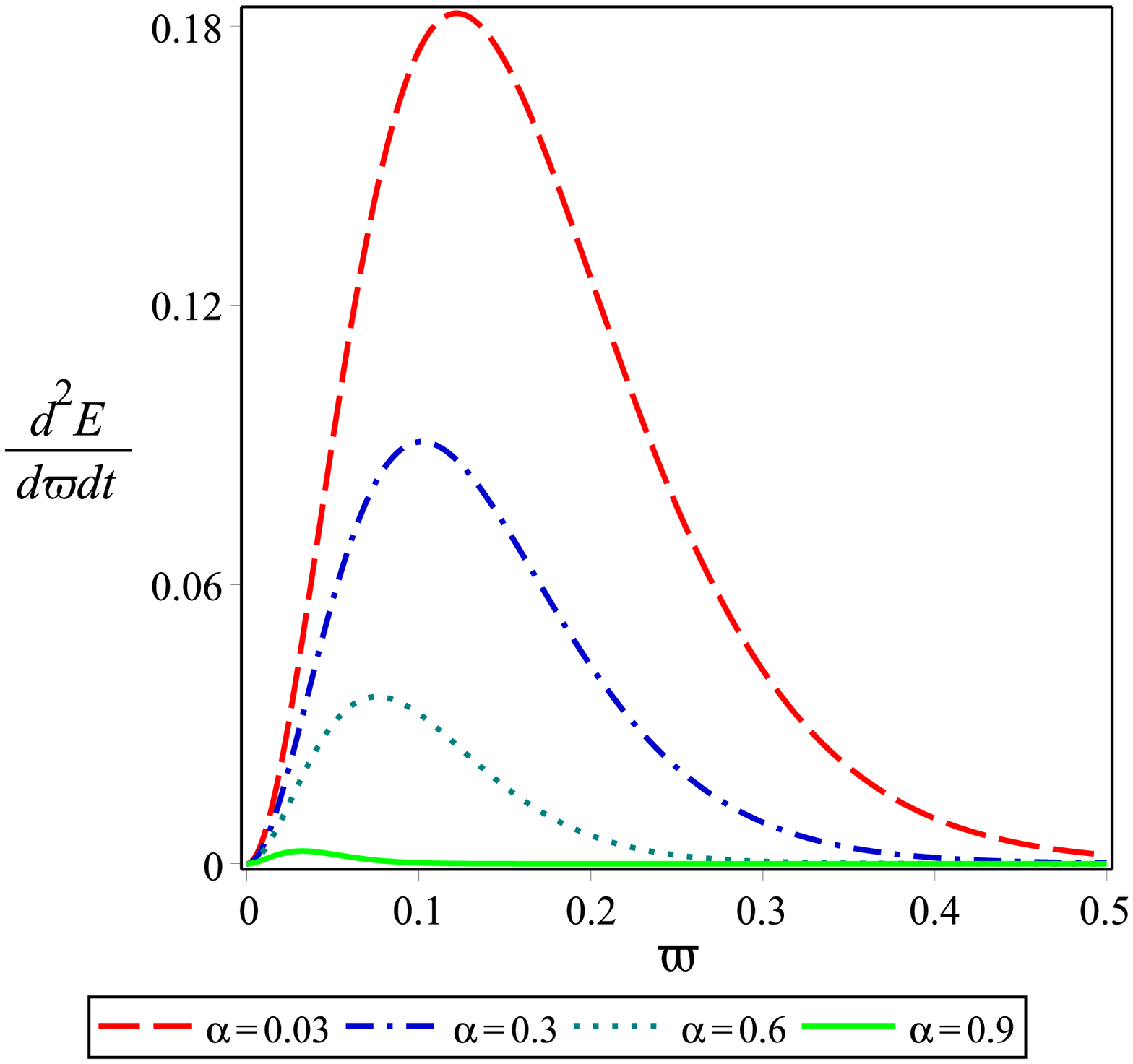}}
\subfloat[$ Q=\alpha=0.2 $ ]{
        \includegraphics[width=0.32\textwidth]{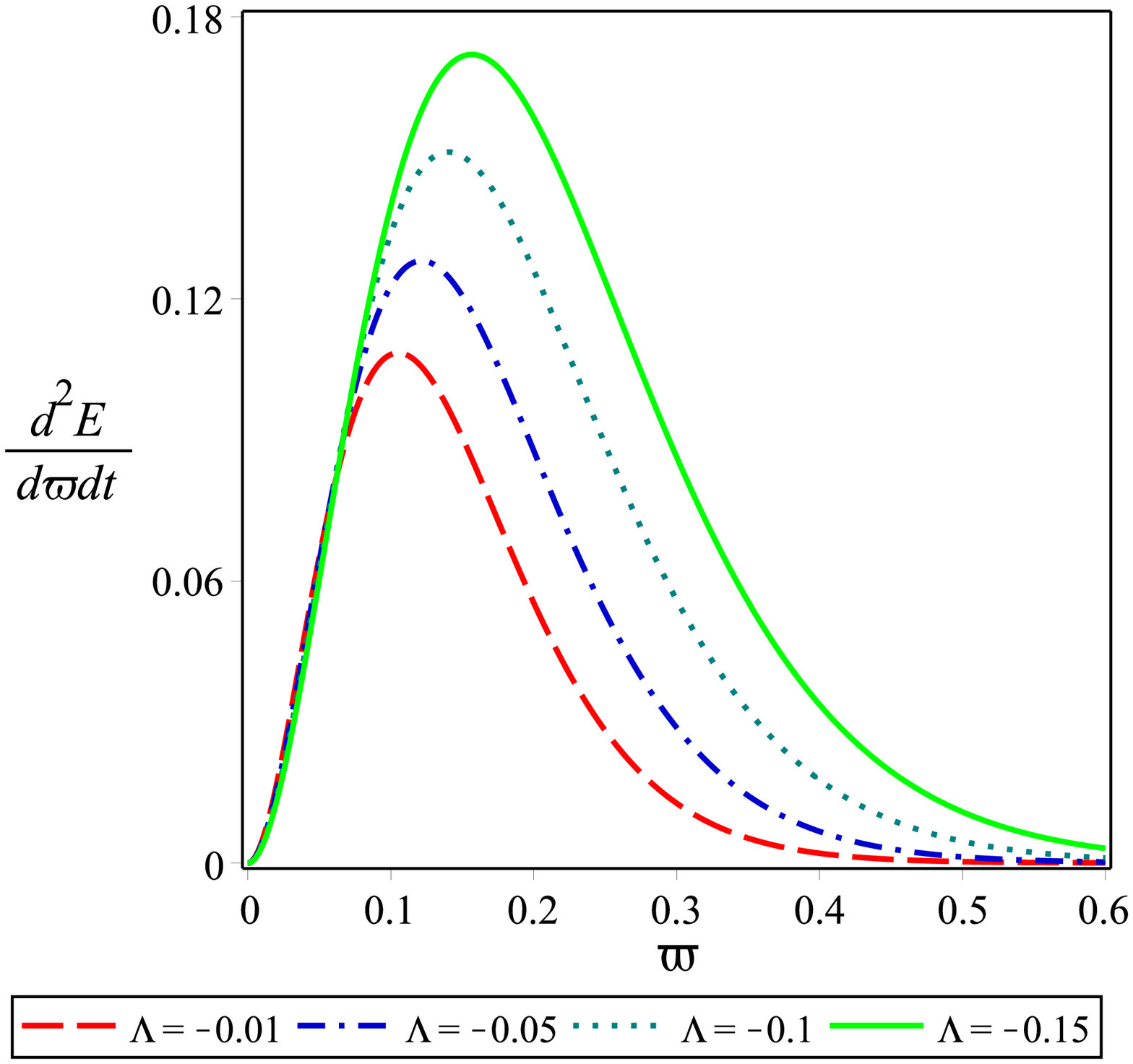}}\newline
\caption{Energy emission rate for the corresponding black hole for $M=1$ and
different values of $Q$, $\protect\alpha $ and $\Lambda $.}
\label{Fig2}
\end{figure}

The energy emission rate is illustrated in Fig. \ref{Fig2} as a function of $%
\varpi $ for different values of the electric charge (left panel), GB
parameter (middle panel) and the cosmological constant (right panel). From
this figure, one can see that there exists a peak of the energy emission
rate for the black hole. When these parameters increase, the peak decreases
and shifts to the low frequency. Since the cosmological constant is
proportional to AdS radius which is representing the natural curvature of
the spacetime, one can find that the evaporation process is slow for a black
hole located in a high curvature background. Similar explanation can be
employed regarding the electric charge and GB parameter. In other words, the
evaporation process would be slow for a black hole with stronger coupling,
or a black hole located in a powerful electric field. As a result, the black
hole has a long lifetime under such conditions. Comparing Fig. \ref{Fig2}b
with Figs. \ref{Fig2}a and \ref{Fig2}c, one can notice that the variation of
$\alpha $ has a stronger effect on the emission of particles around the
black hole. 

\subsection{Deflection angle}

Here, we proceed to study the deflection angle of light by using the null
geodesics method \cite{Chandrasekhar,Weinberg,Kocherlakota,WJaved}. The
total deflection angle can be obtained by the following relation
\begin{equation}
\Theta =2\int_{b}^{\infty }\Big\vert\frac{d\phi }{dr}\Big\vert dr-\pi ,
\label{EqDAn}
\end{equation}%
where $b$ is the impact parameter, defined as $b\equiv \frac{L}{E}$. Using
Eq. (\ref{EqTp}), one can calculate the deflection angle as
\begin{equation}
\Theta =\frac{\alpha }{b^{4}}\left( \frac{Q^{4}}{9b^{2}}-\frac{MQ^{2}}{2b}+%
\frac{4M^{2}}{7}\right) +\left( \frac{Q^{2}}{15b^{2}}-\frac{M}{6b}\right)
\left( 3-2\alpha \Lambda \right) +\frac{7}{3}+(\frac{1}{3}\alpha \Lambda -1)%
\frac{1}{3}\Lambda b^{2}.  \label{EqDAn1}
\end{equation}

\begin{figure}[tbh]
\centering
\subfloat[ $ \alpha=0.2 $ and $ \Lambda =-0.02 $]{
        \includegraphics[width=0.32\textwidth]{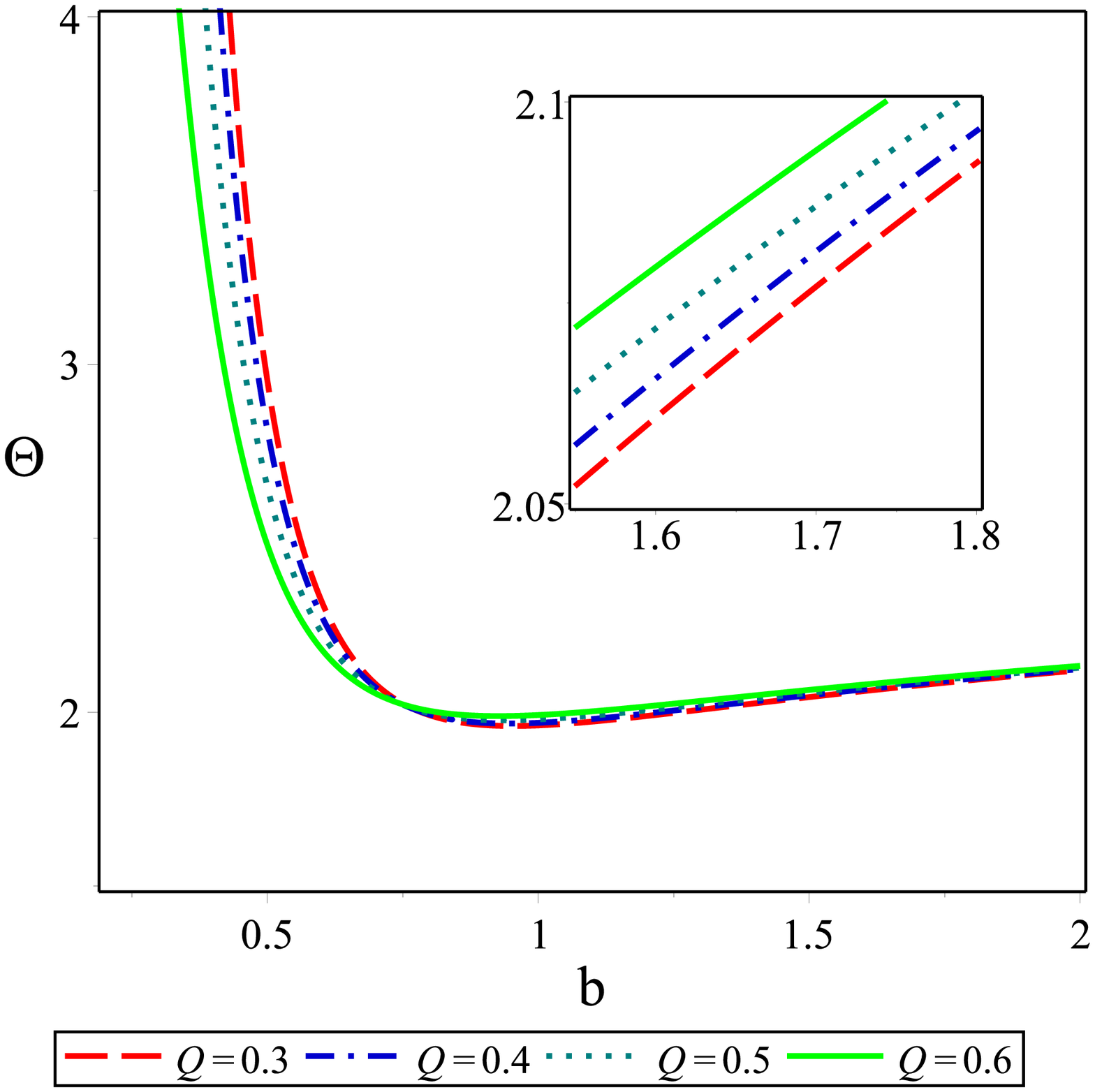}}
\subfloat[$Q=0.2$ and $ \Lambda =-0.02 $]{
        \includegraphics[width=0.32\textwidth]{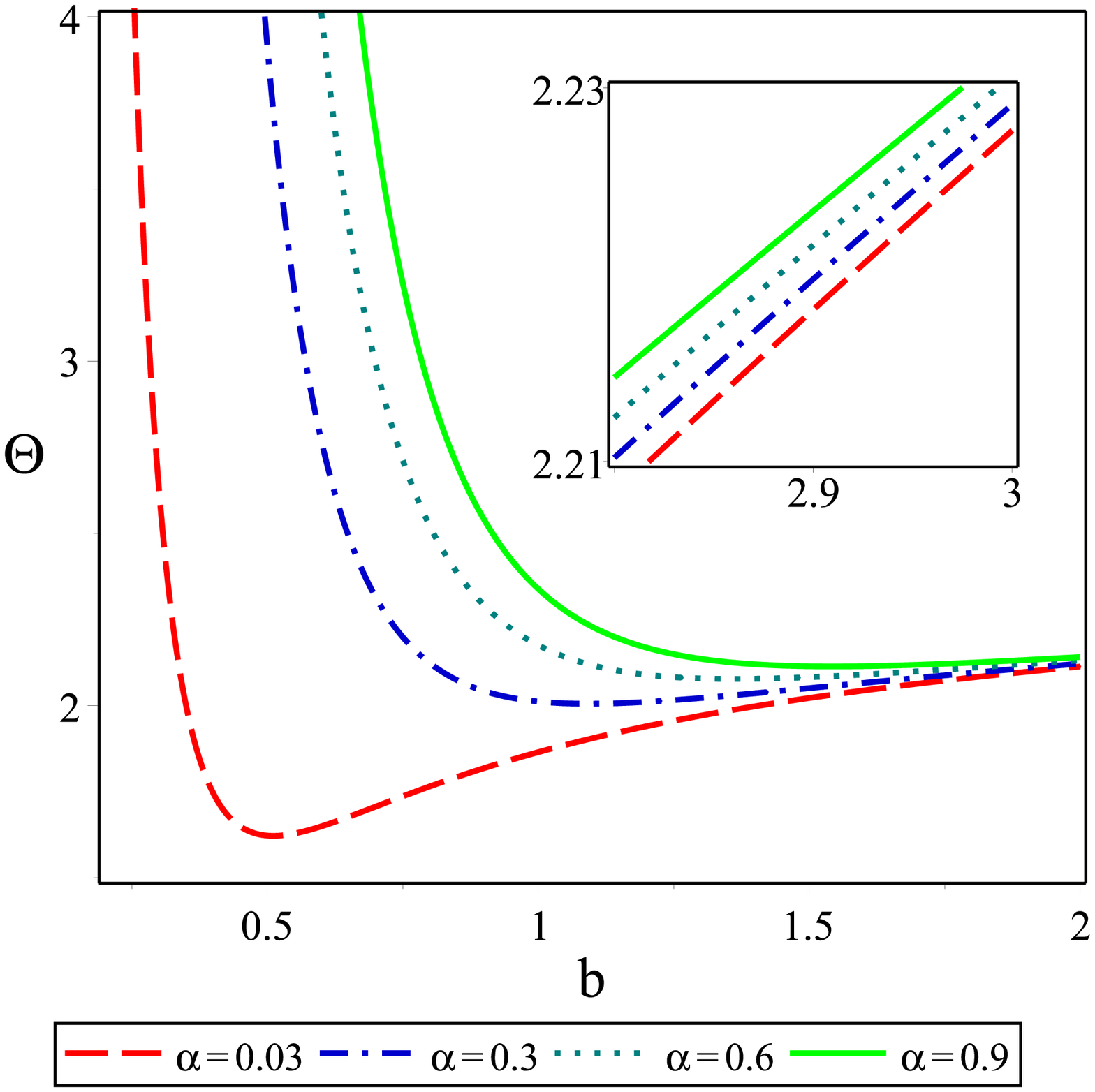}}
\subfloat[$ Q=\alpha=0.2 $ ]{
        \includegraphics[width=0.32\textwidth]{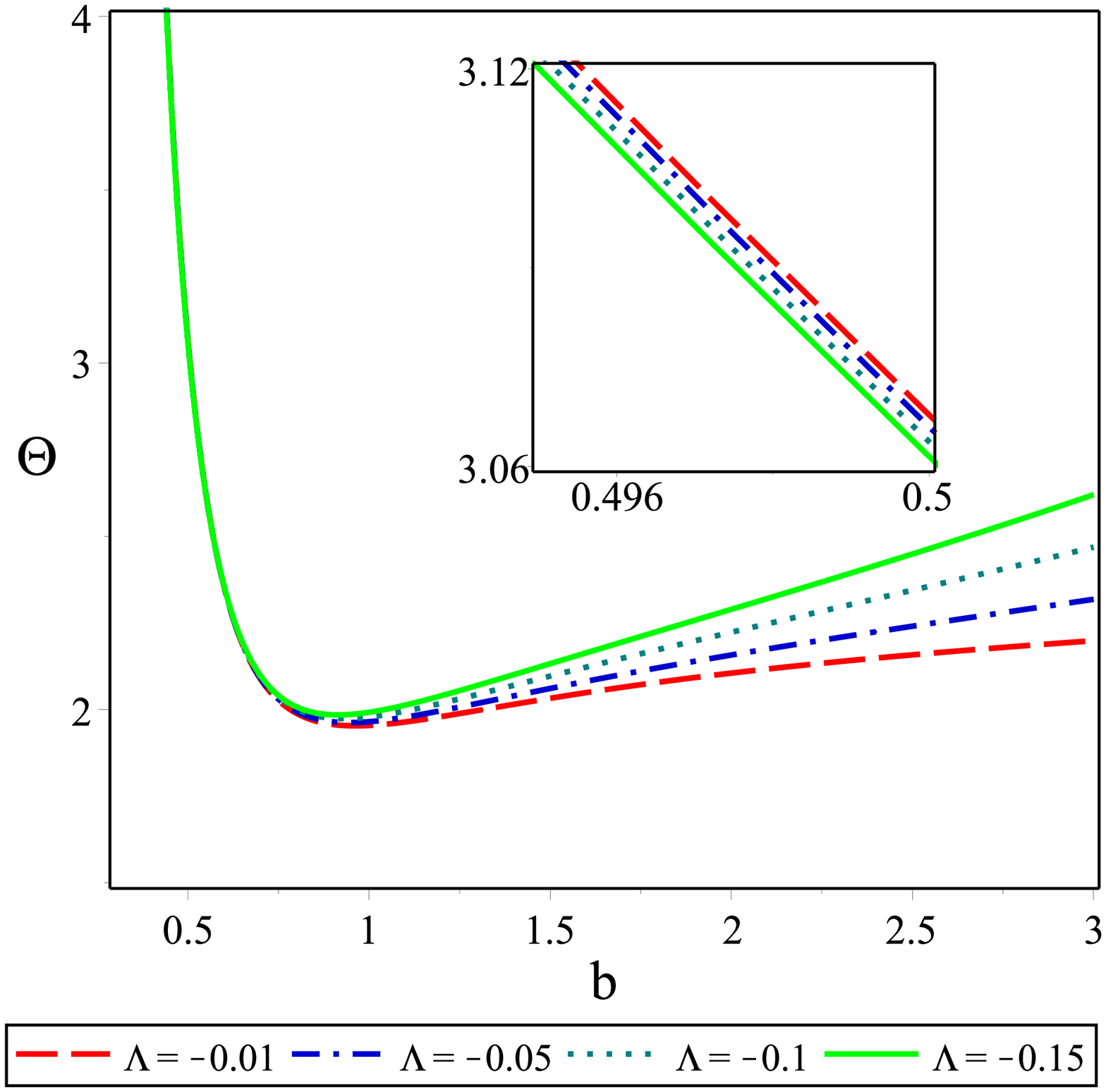}}\newline
\caption{ the behavior of $\Theta $ with respect to impact parameter $b$ for
$M=1$ and different values of $Q$, $\protect\alpha $ and $\Lambda $.}
\label{Fig3}
\end{figure}

To show the effects of different parameters on the deflection angle, we have
plotted Fig. \ref{Fig3}. In this figure, we represent the variation of the
deflection angle $\Theta $ as a function of the parameter $b$ for different
values of $Q$ (left panel), $\alpha $ (middle panel) and $\Lambda $ (right
panel). We can see that all curves reduce rapidly firstly with the growth of
$b$, then they gradually increase by increasing this parameter. In other
words, they have a global minimum value. This means that there exists a
finite value of the impact parameter $b$ which deflection of light is very
low for it. According to the relation $b\equiv \frac{L}{E}$, this finite
value depends on the energy and angular momentum of the photon. In Fig. \ref%
{Fig3}a, we investigate the impact of electric charge on $\Theta $ for fixed
GB parameter and the cosmological constant. We observe that depending on the
value of impact parameter $b$, increasing $Q$ leads to the increasing or
decreasing of the deflection angle. For small values of $b$, increasing $Q$
makes the decreasing of deflection angle, whereas opposite behavior is
observed for large values. Fig. \ref{Fig3}b shows the behavior of $\Theta $
with $b$ for fixed $Q$ and $\Lambda $ and varying $\alpha $. As we see, $%
\Theta $ is an increasing function of $\alpha $ for all values of $b$. To
study the effect of $\Lambda $, we have plotted Fig. \ref{Fig3}c. Taking a
look at this figure, one can find that the cosmological constant has an
increasing (a decreasing) contribution on $\Theta $ for small (large) $b$.
Fig. \ref{Fig3} also displays that for large impact parameters, the effect
of electric charge and GB parameter on the deflection angle is negligible,
whereas the effect of cosmological constant will be notable here. According
to the relation $b\equiv \frac{L}{E}$, one can say that when the energy of a
moving photon gets smaller in comparison to its angular momentum, the effects
of $Q$ and $\alpha $ become weak.

\section{Heat engine}

The discovery of a profound connection between the laws of black hole
mechanics with the corresponding laws of ordinary thermodynamic systems has
been one of the remarkable achievements of theoretical physics \cite%
{Bardeen,Hawking}. In fact, the consideration of a black hole as a
thermodynamic system with a physical temperature and an entropy provides a
deep insight to understand its microscopic structure. In the past two
decades, the study of black hole thermodynamics in an anti-de Sitter (AdS)
space attracted significant attention. Although in early studies, the
cosmological constant was considered as a fixed parameter, the improvement
in the context of black hole thermodynamics showed that the correspondence
between ordinary thermodynamic systems and black hole mechanics would be
completed to include a variable cosmological constant \cite%
{Kubiznak1,Kubiznak2}. Once the variation of $\Lambda $ is included in the
first law, the black hole mass $M$ is identified with enthalpy rather than
internal energy \cite{DKastor}, the cosmological constant is treated as a
thermodynamic pressure $P$ and its conjugate quantity as a thermodynamic
volume $V$ \cite{Dolan}.

\begin{figure*}[tbh]
\centering
\includegraphics[width=0.3\linewidth]{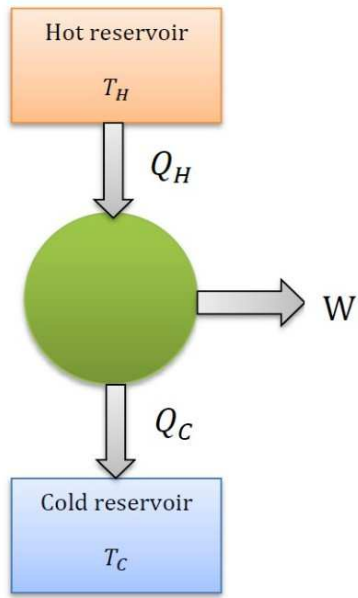}
\caption{The heat engine flows.}
\label{Fig4}
\end{figure*}

Considering black holes as thermodynamic systems in the extended phase
space, it is natural to assume them as heat engines. Indeed, the mechanical
term $PdV$ in the first law provides the possibility of calculating the
mechanical work and in result the efficiency of these heat engines. A heat
engine is defined as a closed path in the $P-V$ diagram which works between
two reservoirs with temperatures $T_{H}$ (high temperature) and $T_{C}$ (low
temperature). During the working process, the heat engine absorbs an amount
of heat $Q_{H}$ from the warm reservoir. Some of this thermal energy is
converted into work $\left( W\right) $ and some heat $\left( Q_{C}\right) $
is usually returned to the cold reservoir (see Fig. \ref{Fig4} for more
details). The efficiency of the heat engine is defined by
\begin{equation}
\eta =\frac{W}{Q_{H}}=1-\frac{Q_{C}}{Q_{H}}.  \label{Eqeff}
\end{equation}

\begin{figure*}[tbh]
\centering
\includegraphics[width=0.38\linewidth]{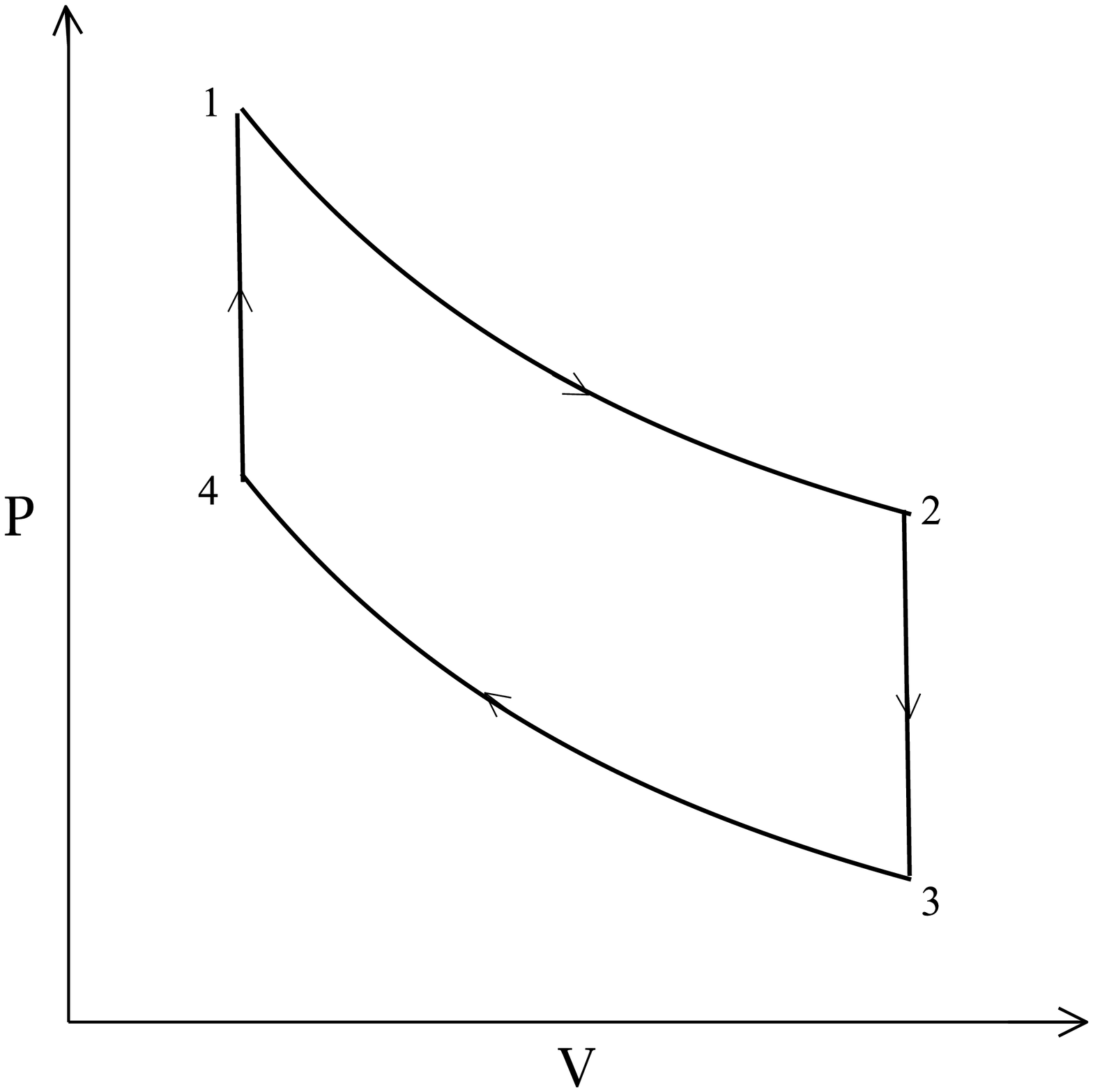}\hfil
\includegraphics[width=0.38\linewidth]{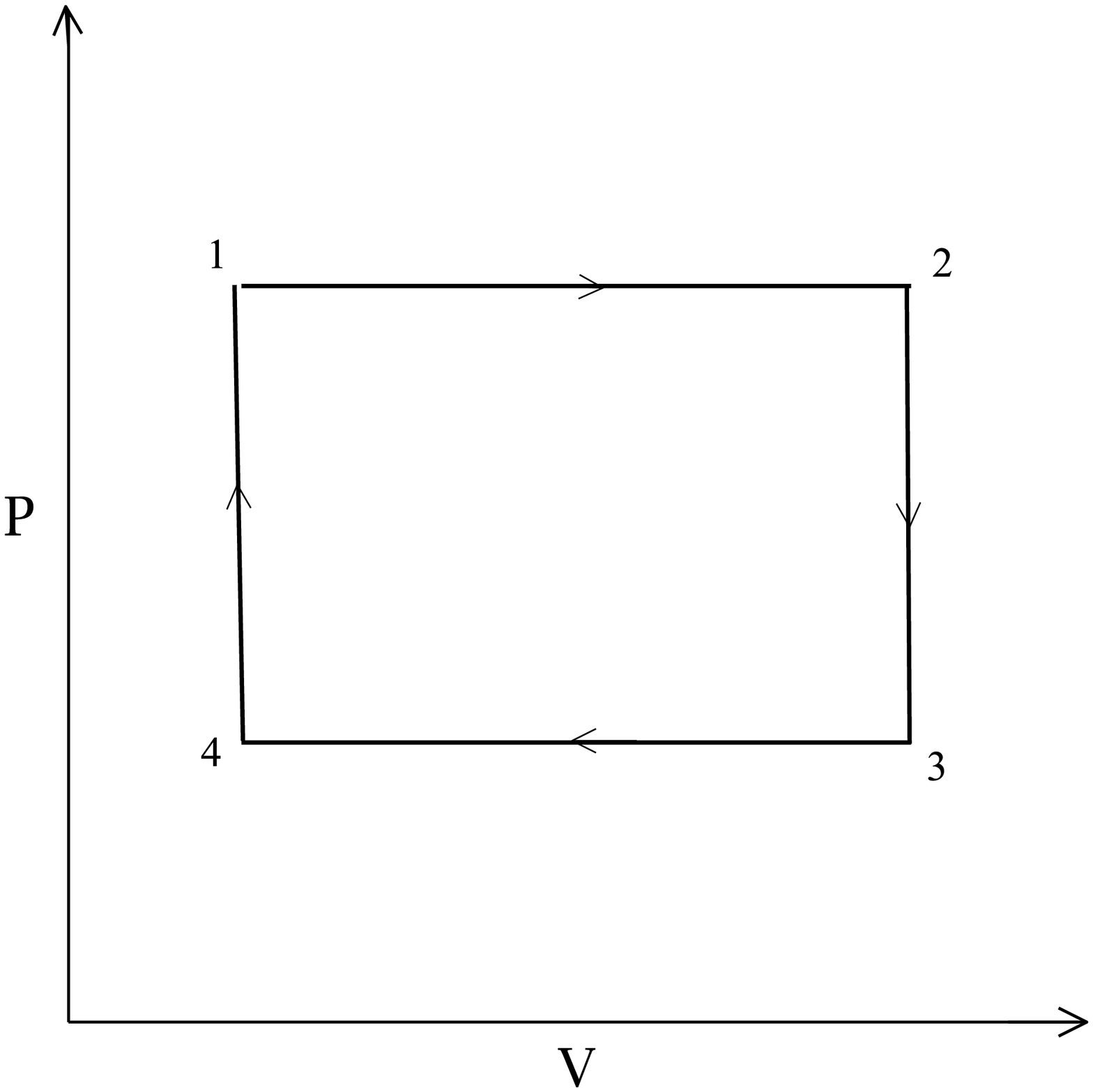}
\caption{$P$-$V$ diagram of thermodynamic cycles for the heat engine: Left
panel: Carnot engine, which for static black holes is also a Stirling
engine. Right panel: the rectangle cycle.}
\label{Fig5}
\end{figure*}

The heat engine efficiency depends on the equation of state of the black
hole and the paths forming the heat cycle in the $P-V$ diagram. There are
different classical cycles which the Carnot cycle is the simplest cycle that
can be considered. This cycle involves a pair of isotherms at temperatures $%
T_{H}$ and $T_{C}$ where the system absorbs some heat during an isothermal
expansion and loses some of this thermal energy during an isothermal
compression. These two temperatures are connected with adiabatic paths. The
Carnot efficiency is determined as
\begin{equation}
\eta _{c}=1-\frac{T_{C}}{T_{H}},
\end{equation}%
this the maximum efficiency any heat engine can have and any higher
efficiency would violate the second law of thermodynamics. For the static
black holes, the thermodynamic volume $V$ and the entropy $S$ are only a
function of event horizon $r_{+}$. In fact, they are dependent on each
other. So, heat capacity equals to zero at constant volume $(C_{V}=0)$. The
vanishing of $C_{V}$ results into "isochore equals adiabatic". In other
words, Carnot and Stirling cycles coincide (see left panel of Fig. \ref{Fig5}%
). Therefore an explicit expression for $C_{P}$ would suggest that one can
define a new engine cycle in $P-V$ plane, as a rectangle (see right panel of
Fig. \ref{Fig5}) which is composed of two isobars (paths of $1\rightarrow 2$
and $3\rightarrow 4$) and two isochores (paths of $2\rightarrow 3$ and $%
4\rightarrow 1$). We can calculate the work done along the heat cycle as
\begin{eqnarray}
W &=&\oint PdV=W_{1\longrightarrow 2}+W_{2\longrightarrow
3}+W_{3\longrightarrow 4}+W_{4\longrightarrow 1}  \notag \\
&&  \notag \\
&=&W_{1\longrightarrow 2}+W_{3\longrightarrow 4}=P_{1}\left(
V_{2}-V_{1}\right) +P_{4}\left( V_{4}-V_{3}\right) ,  \label{EqWo}
\end{eqnarray}%
in the above equation, the work done along paths of $2\rightarrow 3$ and $%
4\rightarrow 1$ are zero ($W_{2\longrightarrow 3}=W_{4\longrightarrow 1}=0$).

The upper isobar will give the heat input as
\begin{eqnarray}
Q_{H} &=&\int_{T_{1}}^{T_{2}}C_{P}\left( P_{1},T\right)
dT=\int_{S_{1}}^{S_{2}}C_{P}\left( P_{1},T\right) \left( \frac{\partial T}{%
\partial S}\right) dS  \notag \\
&&  \notag \\
&=&\int_{S_{1}}^{S_{2}}TdS=\int_{H_{1}}^{H_{2}}=M_{2}-M_{1}.  \label{EqQH}
\end{eqnarray}

In 2014, Johnson in Ref. \cite{Johnson}, calculated the efficiency of heat
engine for a black hole. Using the concepts introduced by Johnson, the
efficiency of heat engines for other types of black holes such as GB black
holes \cite{JohnsonII}, Born--Infeld AdS black holes \cite{JohnsonIII}, dilatonic
Born-Infeld black holes \cite{Bhamidipati}, rotating black holes \cite%
{Hennigar}, charged AdS black holes \cite{Liu}, Kerr AdS and dyonic black
holes \cite{Sadeghi}, BTZ black holes \cite{Mo}, polytropic black holes \cite{Setare},
AdS black holes in higher dimensions \cite{BelhajC}, black holes in
conformal gravity \cite{Xu}, black holes in massive gravity \cite{Hendi},
benchmarking black holes \cite{Chakraborty}, accelerating AdS black holes
\cite{Zhang}, black holes in gravity's rainbow \cite{EslamPanah}, charged
accelerating AdS black holes \cite{ZhangII}, nonlinear charged AdS black
holes \cite{Nam}, charged rotating accelerating AdS black holes \cite%
{Jafarzade}, and Hayward-AdS black holes \cite{SGuo}.

\subsection{Charged EGB-AdS black hole as holographic heat engine}

Now, we want to study the charged AdS black hole in $4D$ EGB gravity as a 
holographic heat engine. The mass of black hole is obtained by solving  the metric function on horizon $(f(r=r_{e})=0)$ in the following form,
\begin{equation}
M=\frac{r_{e}}{2}\left( 1-\frac{\Lambda r_{e}^{2}}{3}+\frac{Q^{2}+\alpha }{%
r_{e}^{2}}\right) .  \label{Mass}
\end{equation}

Using Eqs. (\ref{EqT-H}) and (\ref{Mass}), the black hole entropy is given
by,
\begin{equation}
S=\int_{0}^{r_{e}}\frac{dM}{T}=\pi r_{e}^{2}+4\pi \alpha \ln \left( \frac{%
r_{e}}{l_{0}}\right) ,
\end{equation}%
where $l_{0}$ is an arbitrary constant with length dimension which is coming
from the fact that the logarithmic arguments should be dimensionless.

In the extended phase space, the cosmological constant corresponds to
thermodynamic pressure with $\Lambda =-8\pi P$, and its conjugate variable
corresponds to thermodynamic volume with

\begin{equation}
V=\left( \frac{\partial M}{\partial P}\right) _{S,Q,\alpha }=\frac{4}{3}\pi
r_{e}^{3}.
\end{equation}

Now, we are going to investigate holographic heat engine for this solution.
Using Eq. (\ref{EqWo}), the useful work is obtained as
\begin{equation}
W=\frac{4}{3}\pi l_{0}^{3} \left( P_{1}-P_{4}\right) \exp \left( -\frac{6\pi
\alpha LambertW\left( \frac{ l_{0}^{2}e^{S/2\pi \alpha }}{2\alpha }\right)
-3S}{4\pi \alpha }\right) \Bigg\vert_{S_{1}}^{S_{2}},  \label{work}
\end{equation}
and $Q_{H}$ is calculated as

\begin{equation}
Q_{H}=\frac{\chi _{1}}{6}\left( 3+8\pi P_{1}\chi _{2}+\frac{3\left( \alpha
+Q^{2}\right) }{\chi _{2}}\right) \Bigg\vert_{S_{1}}^{S_{2}},  \label{QH}
\end{equation}%
where%
\begin{equation}
\chi _{n}=l_{0}^{n}\exp \left( -\frac{2n\pi \alpha LambertW\left( \frac{%
l_{0}^{2}e^{S/2\pi \alpha }}{2\alpha }\right) -nS}{4\pi \alpha }\right) .
\end{equation}

Inserting Eqs. (\ref{work}) and (\ref{QH}) into Eq. (\ref{Eqeff}), one can
obtain the engine efficiency. To calculate Carnot efficiency, we consider
the $T_{H}$ and $T_{C}$ in our cycle correspond to $T_{2}$ and $T_{4}$,
respectively. So, this efficiency is

\begin{eqnarray}
\eta _{c} &=&1-\frac{T\left( P_{4},S_{1}\right) }{T\left( P_{1},S_{2}\right)
}  \notag \\
&&  \notag \\
&=&1-\frac{y_{1}\left( y_{2}+2\alpha \right) \left( 8\pi
P_{4}x_{4}+x_{2}-Q^{2}-\alpha \right) }{x_{1}\left( x_{2}+2\alpha \right)
\left( 8\pi P_{4}y_{4}+y_{2}-Q^{2}-\alpha \right) },  \label{QH1}
\end{eqnarray}%
where
\begin{eqnarray}
x_{n} &=& l_{0}^{n}\exp \left( -\frac{2n\pi \alpha ~LambertW\left( \frac{%
l_{0}^{2}e^{S/2\pi \alpha }}{2\alpha }\right) -nS_{1}}{4\pi \alpha }\right) ,
\notag \\
&&  \notag \\
y_{n} &=& l_{0}^{n}\exp \left( -\frac{2n\pi \alpha ~LambertW\left( \frac{%
l_{0}^{2}e^{S/2\pi \alpha }}{2\alpha }\right) -nS_{2}}{4\pi \alpha }\right) .
\end{eqnarray}

\begin{figure*}[tbh]
\centering
\includegraphics[width=0.38\linewidth]{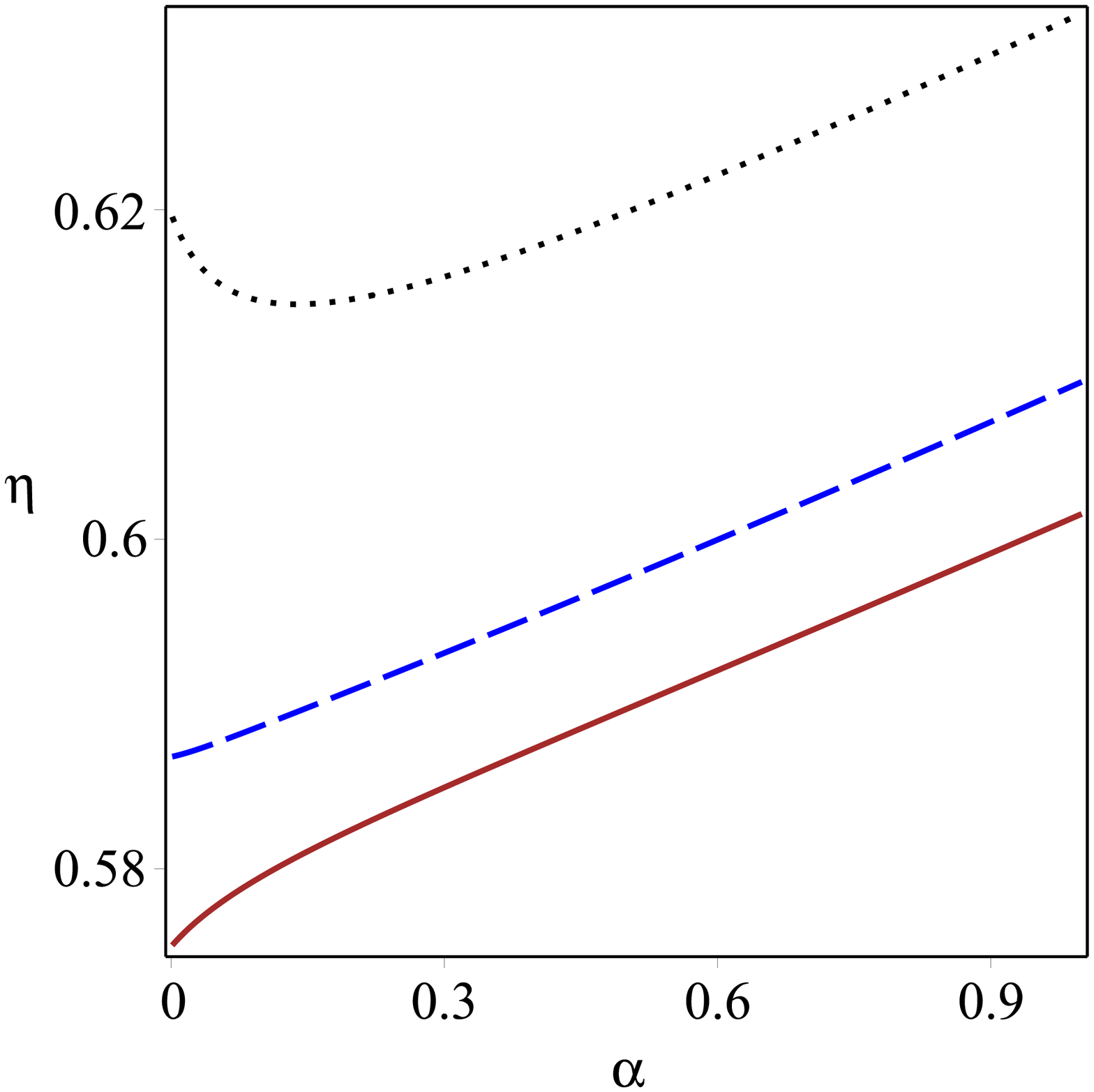}\hfil
\includegraphics[width=0.38\linewidth]{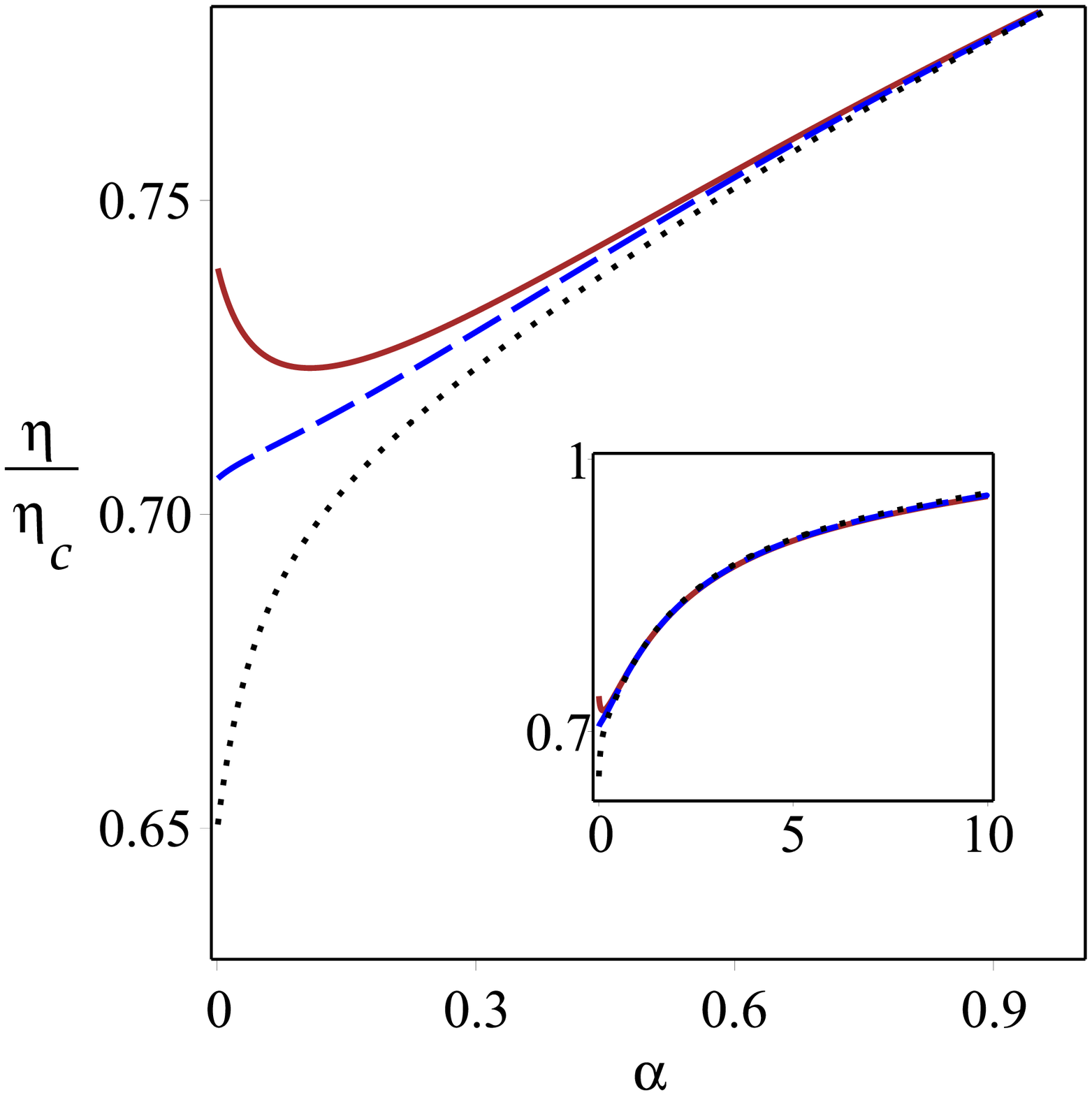}
\caption{Variation of $\protect\eta $ and $\frac{\protect\eta}{\protect\eta %
_{C}}$ versus $\protect\alpha$ for $P_{1}=1$, $P_{4}=0.4$, $l_{0}=1$, $S_{2}=5$, $
S_{1}=1$,  $Q=0.2$ (continuous line), $Q=0.6$ (dashed line) and $Q=1$ (dotted
line).}
\label{Fig6}
\end{figure*}

\begin{figure}[!htb]
\centering
\includegraphics[width=0.38\textwidth]{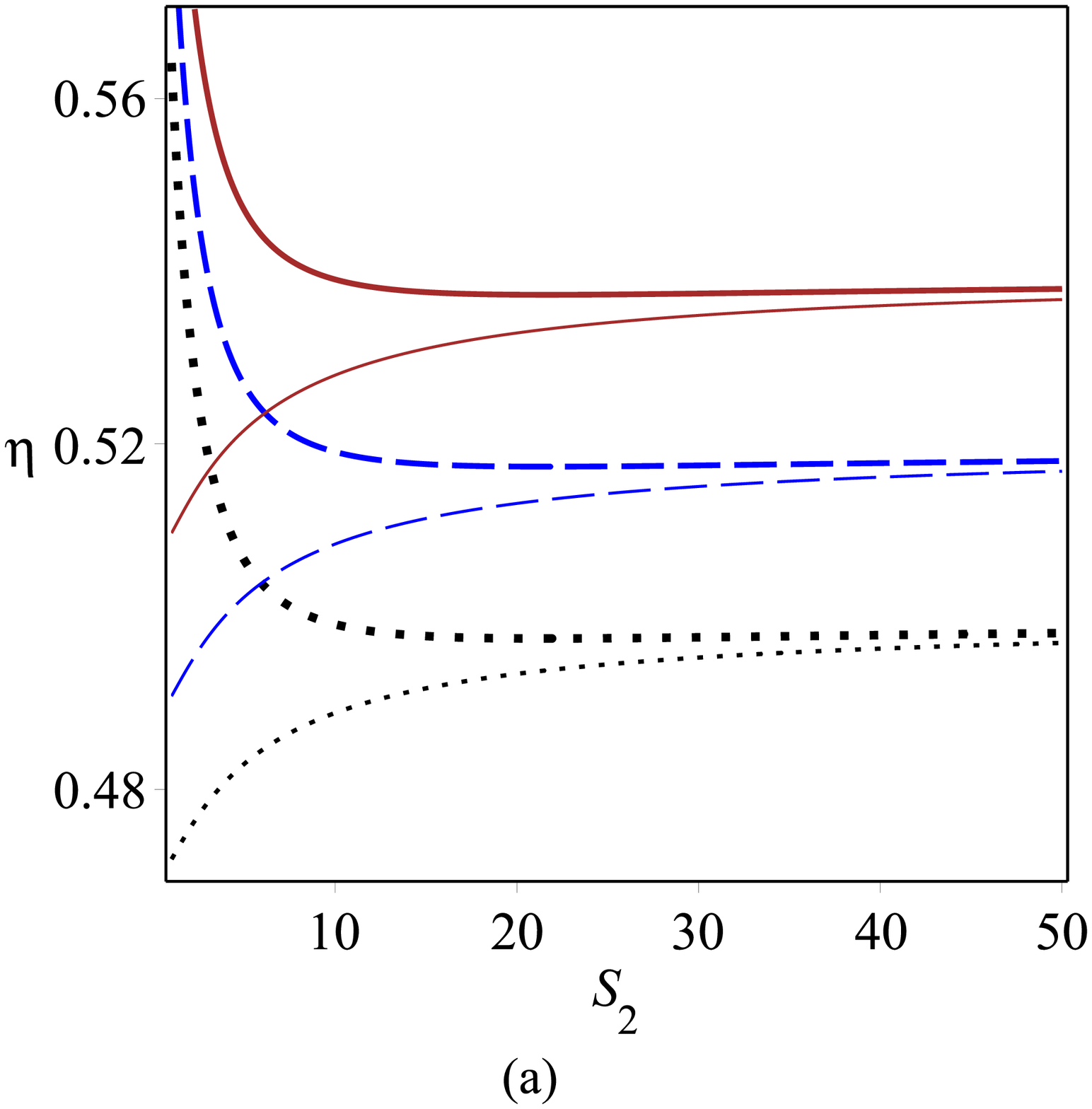}
\includegraphics[width=0.38
\textwidth]{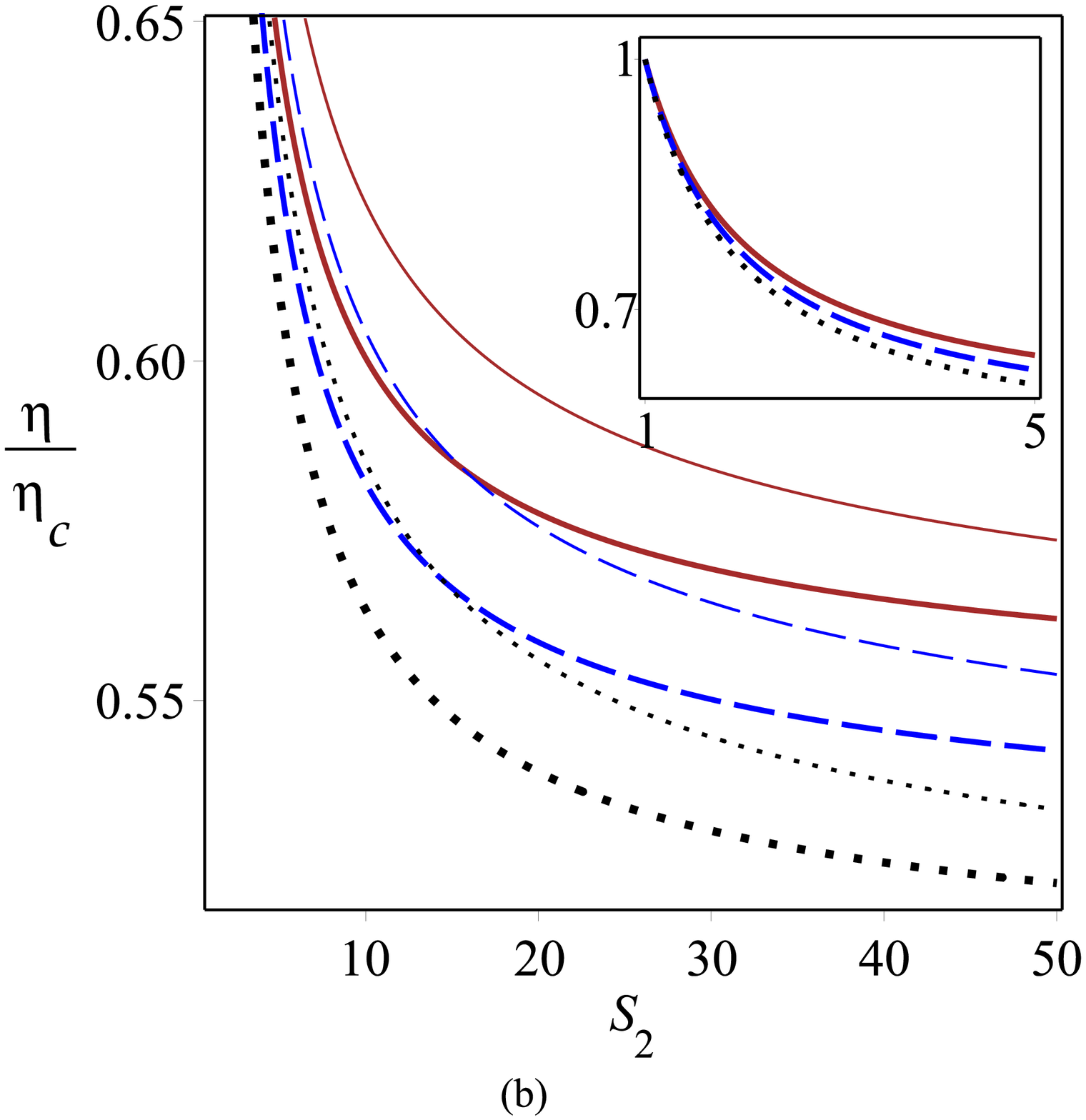} \newline
\includegraphics[width=0.38\textwidth]{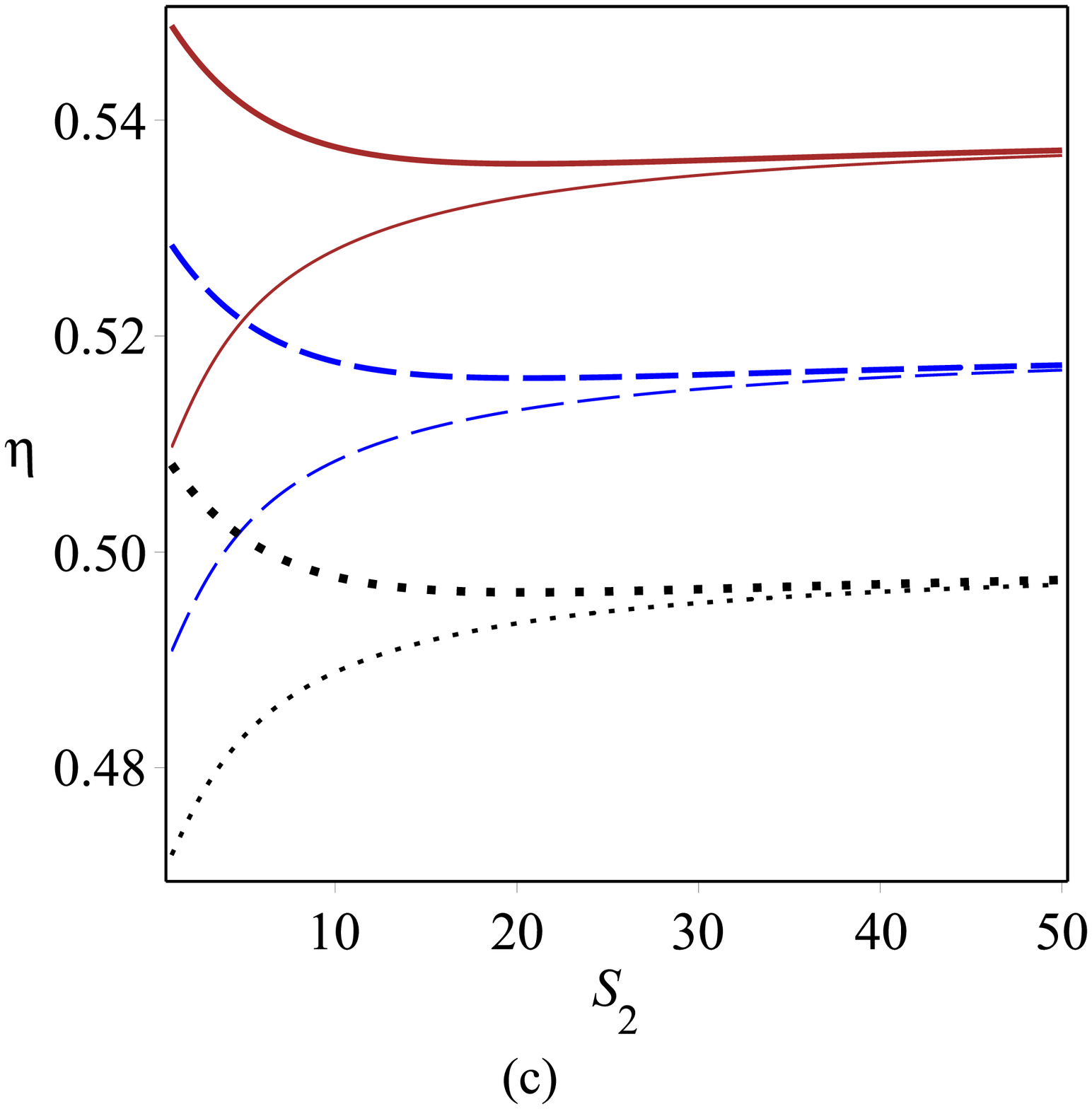}
\includegraphics[width=0.38
\textwidth]{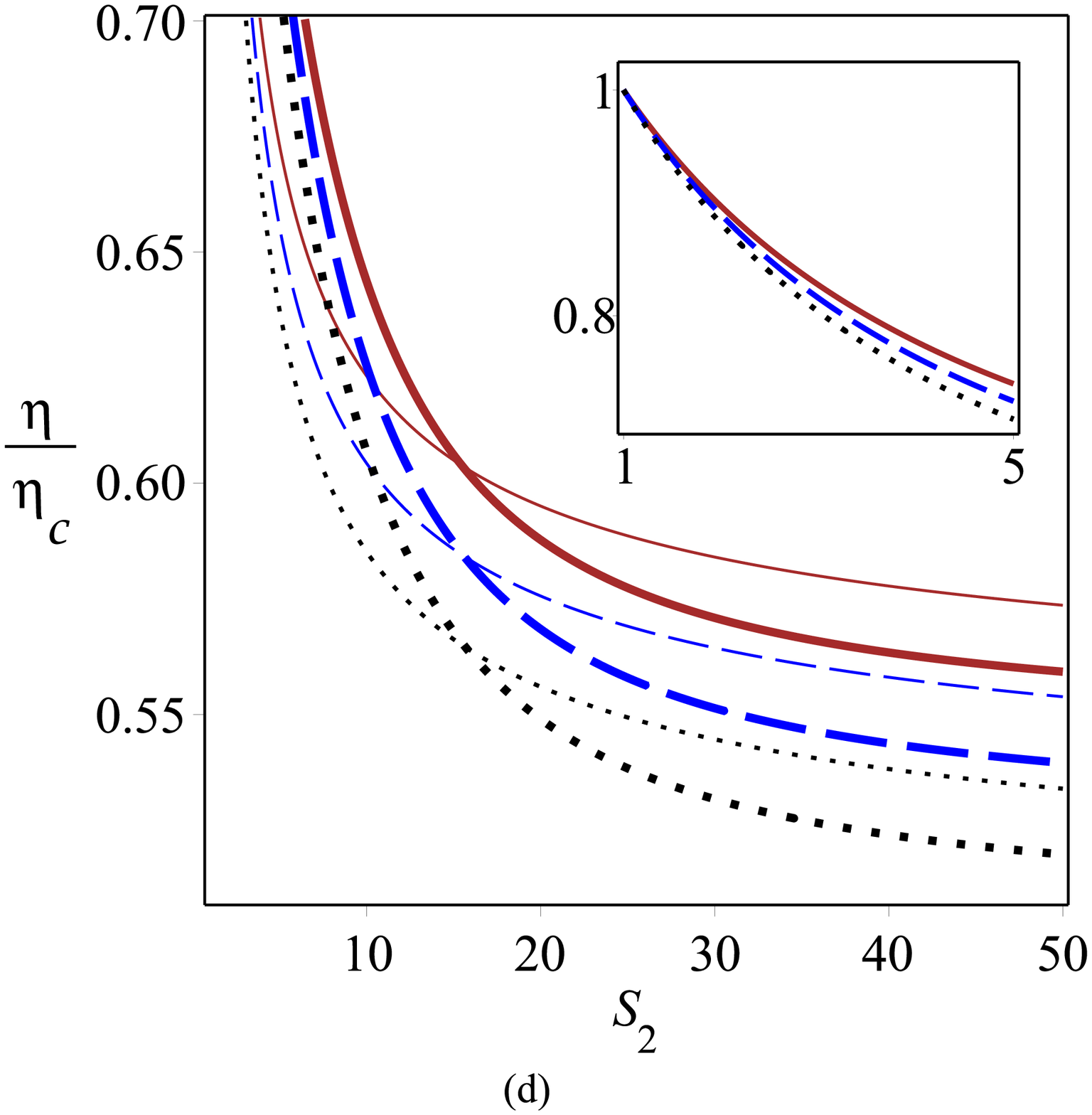} \newline
\caption{Variation of $\protect\eta $ and $\frac{\protect\eta}{\protect\eta 
_{C}}$ versus $S_{2}$ for $S_{1}=l_{0}=1$, $\Delta P=0.54$ (continuous line), $
\Delta P=0.52$ (dashed line) and $\Delta P=0.5$ (dotted line). Up panels for
$\protect\alpha=0.1 $, $Q=1$ (bold lines) and $Q=0.2 $ (thin lines). Down
panels for $Q=0.2 $, $\protect\alpha=1 $ (bold lines) and $\protect\alpha %
=0.1 $ (thin lines).}
\label{Fig7}
\end{figure}

In order to study the effects of electric charge and GB coupling on the heat
engine efficiency, we have plotted Fig. \ref{Fig6}. The left panel of Fig. %
\ref{Fig6}, shows variation of efficiency $\eta $ versus GB parameter for
different values of the electric charge with fixed pressure $\left(
P_{1},P_{4}\right) $ and entropy $\left( S_{1},S_{2}\right) $. From this
figure, one can observe that the behavior of the efficiency is crucially
dependent on the electric charge and GB coupling. As we see, $\eta $ is an
increasing function of the electric charge. For small values of $Q$, the
efficiency is a monotonously increasing function with the growth of $\alpha $%
. Whereas, for large $Q$, the efficiency curve has a global minimum value.
This reveals the fact that there is a certain value of the GB parameter at
which the heat engine of the black hole works at the lowest efficiency. The
right panel of Fig. \ref{Fig6}, exhibits the ratio between the efficiency
and the Carnot efficiency $\left( \frac{\eta }{\eta _{c}}\right) $ versus $%
\alpha $ for different values of $Q$. From this figure, it is evident that
the efficiency is close to the Carnot efficiency for large values of $\alpha
$. For large electric charges, the ratio $\frac{\eta }{\eta _{c}}$ monotonic
increases as the GB parameter increases, whereas for small values this ratio
decreases rapidly firstly and reaches a minimum value, then the efficiency
grows and approaches the Carnot efficiency by increasing $\alpha $. This
figure also shows that the condition $\frac{\eta}{\eta_{c}} < 1$ always
holds and this result is consistent with the second law of the
thermodynamics (Carnot heat engine has the highest efficiency). In order to
see the effect of pressure on the heat engine efficiency, we have plotted $%
\eta $ versus entropy $S_{2}$ for different values of pressure difference $%
\Delta P$ with fixed $Q$ and $\alpha $ in Fig. \ref{Fig7}. We observe that
increasing the pressure makes the increase in efficiency (see Figs. \ref%
{Fig7}a and \ref{Fig7}c). As we see, depending on the values of the GB
parameter and electric charge, increasing entropy $S_{2}$ leads to
increasing or decreasing of the heat engine efficiency. By looking at bold
lines in two left panels of Fig. \ref{Fig7}, one can find that for black
holes with a large electric charge or stronger GB coupling, the efficiency
decreases with the growth of the entropy $S_{2}$ (corresponding to volume $%
V_{2}$). This shows that the heat engine efficiency gets smaller when volume
difference between the small black hole $\left( V_{1}\right) $ and large
black hole $\left( V_{2}\right) $ becomes bigger. Whereas, opposite behavior
will be observed for a small electric charge and weak GB coupling constant (see thin
lines of Figs. \ref{Fig7}a and \ref{Fig7}c). From two right panels of Fig. 
\ref{Fig7}, we see that the ratio $\frac{\eta }{\eta _{c}}$ is a
monotonously decreasing function of the entropy $S_{2}$. In the limit of
that the entropy $S_{2}$ goes to the infinity, this ratio reaches a constant
value. Also, these two figures show that the effciency is close to the
Carnot effciency for small volume difference $(\Delta V=V_{2} -V_{1})$ and
it will never be bigger than Carnot heat engine, this is consistent to the
heat engine nature in traditional thermodynamics.


\section{CONCLUSION}

Among the higher curvature gravity theories, the most extensively studied
theory is the so-called EGB gravity. Recently there has been a particular
interest in regularizing $D-$dimensional EGB solution to obtain a $4D$ EGB
gravity in $D\rightarrow 4$ limit. This theory admits a static spherically
symmetric black hole and possesses only the degrees of freedom of a massless
graviton and thus free from the instabilities. Such an interesting property
motivates one to investigate different properties of the $4D$ EGB black hole
in AdS space.

First, we investigated the photon sphere and the shadow observed by a
distant observer and explore the effects of black hole parameters on them.
The results showed that both the photon sphere radius and the shadow size
decrease with the increasing GB parameter. Studying the impact of electric
charge, we noticed that the shadow size shrinks with the increasing electric
charge which is the same as the results obtained for variation of the photon
sphere radius. We also found that the cosmological constant has an
increasing effect on the radius of shadow and its effect is stronger than
other parameters.

Then, we continued by investigating the energy emission rate and examining
the influence of parameters on the radiation process. The results
illustrated that as the effects of electric charge and coupling constants
get stronger or the curvature of the background becomes higher, the
evaporation process gets slower. In other words, the lifetime of a black
hole would be longer under such conditions.

Also, we studied the gravitational lensing of light around the $4D$ EGB
black holes. Depending on the parameters of the black hole, photons get
deflected from their straight path and have different behaviors. The results
are summarized as: i) the deflection angle was a decreasing (an increasing)
function of $b$ for small (large) values of the impact parameter. ii) the GB
parameter had an increasing effect on the deflection angle, whereas the
effects of electric charge and cosmological constant were dependent on the
values of impact parameter. For small values of $b$, the deflection angle
decreased (increased) with increasing of the electric charge (cosmological
constant). For large values, their effects were the opposite.

Finally, we have considered charged $4D$ EGB-AdS black holes as a working
substance and studied the holographic heat engine by taking a rectangle heat
cycle in the $P-V$ plot. First, we investigated the concept of a heat engine
for the charged EGB-AdS black hole and showed that the electric charge, GB
coupling and thermodynamic pressure affect significantly the efficiency of
black hole heat engine. We observed that efficiency will increase with the
growth of electric charge and pressure. Regarding the effect of GB coupling,
we noticed that in the presence of a weak electric field, the efficiency of
black hole is an increasing function of the GB parameter. Whereas, for black
holes with a stronger electric charge, the efficiency will have a global
minimum for a certain value of this parameter. In other words, there is a
finite value of the GB coupling at which the heat engine of this black hole
works at the lowest efficiency. Also, we found that the heat engine
efficiency will approach Carnot efficiency if the volume difference between
small and large black holes becomes very small. In Addition, we found that
the ratio of $\frac{\eta}{\eta_{c}}$ versus entropy $S_{2}$ is less than one
all the time which is consistent with the thermodynamic second law.

\begin{acknowledgements}
BEP and SHH thank Shiraz University Research Council. The work of
BEP has been supported financially by Research Institute for
Astronomy and Astrophysics of Maragha (RIAAM) under research
project No. 1/6025-30.
\end{acknowledgements}

\end{document}